\documentclass[12pt]{article}
\usepackage{a4wide}
\usepackage{latexsym}
\usepackage{cite}
\usepackage{graphicx}
\usepackage{booktabs}
\usepackage{color}

\usepackage{pslatex}
\usepackage[latin1]{inputenc}
\usepackage[T1]{fontenc}

\def\bq{\begin{eqnarray}}
\def\eq{\end{eqnarray}}

\newcommand{\qbar}{\ensuremath{\bar{q}}}
\newcommand{\pT}[1]{\ensuremath{p_{\perp#1}}}
\newcommand{\mrm}[1]{\ensuremath{\mathrm{#1}}}

\def\lsim{\mathrel{\rlap{\lower4pt\hbox{\hskip1pt$\sim$}}
    \raise1pt\hbox{$<$}}}                
\def\gsim{\mathrel{\rlap{\lower4pt\hbox{\hskip1pt$\sim$}}
    \raise1pt\hbox{$>$}}}                

\newcommand{\plotset}[1]{
\begin{tabular}[c]{c}
\hspace*{-0.7cm}\includegraphics*[scale=0.4]{PSdivA4Avg-#1.eps}\hspace*{-0.8cm} \end{tabular}&
\begin{tabular}[c]{c}\vspace*{3mm}
 \hspace*{-0.2cm}\includegraphics*[scale=0.4]{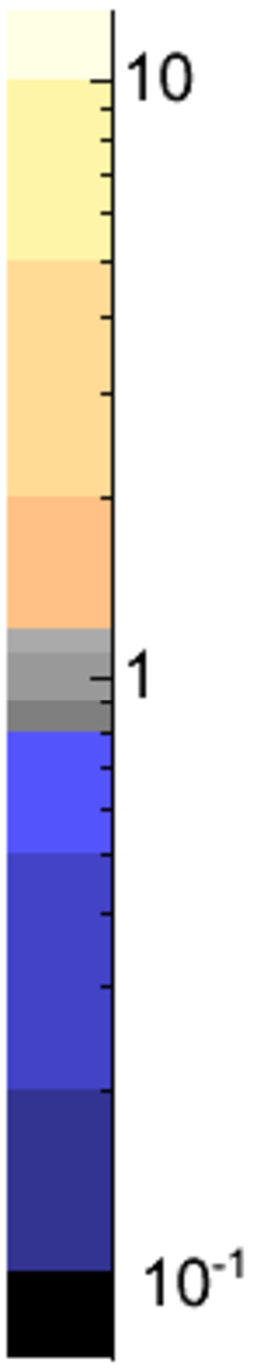}\hspace*{-2.2cm}
\end{tabular}
& 
\begin{tabular}[c]{c}
\hspace*{-0.6cm}\includegraphics*[scale=0.4]{PSdivA4Rms-#1.eps}\hspace*{-1.3cm} \end{tabular}
}

\begin{document}

\thispagestyle{empty}

\begin{flushright}
FERMILAB-PUB-09-035-T\\
MCnet/09/05\\
MZ-TH/09-07 
\end{flushright}

\vspace{1.5cm}

\begin{center}
  {\Large\bf 
   Some remarks on dipole showers and the DGLAP equation\\
  }
  \vspace{1cm}
  {\large 
    Peter Skands ${}^{1}$ and Stefan Weinzierl ${}^{2}$\\
  \vspace{1cm}
      {\small ${}^{1}$ \em Theoretical Physics, Fermilab, MS 106, Box 500,}\\
      {\small \em Batavia, IL - 60510, USA}\\
  \vspace{2mm}
      {\small ${}^{2}$ \em Institut f{\"u}r Physik, Universit{\"a}t Mainz,}\\
      {\small \em D - 55099 Mainz, Germany}\\
  } 
\end{center}

\vspace{2cm}

\begin{abstract}\noindent
  {
It has been argued recently that parton showers
based on colour dipoles conflict with collinear factorization and do not
lead to the correct DGLAP equation.
We show that this conclusion is based on an inappropriate
assumption, namely the choice of the gluon energy as evolution variable.
We further show numerically that Monte Carlo programs based on
dipole showers with ``infrared sensible'' evolution variables 
reproduce the DGLAP equation both in
asymptotic form as well  
as in comparison to the leading behaviour of second-order QCD matrix elements.   }
\end{abstract}

\vspace*{\fill}

\newpage

\section{Introduction}

Due to the large centre-of-mass (CM) energies at modern hadron 
colliders, and especially in
connection with the production of high-$\hat{s}$ states
it is becoming increasingly essential that we have
well-controlled calculations 
that cover all regions of phase space in a reliable way (see, e.g.,
\cite{Alwall:2008qv}), 
and that these are implemented consistently in the phenomenological  
tools we use for collider studies --- we shall here focus on
Monte Carlo Event Generators. It is likewise important that we have an
accurate theoretical understanding of the properties we rely on these
calculations to have. 

Non-trivial collider observables almost invariably involve an interplay
between widely separated energy scales. This represents an important
challenge to collider phenomenology since,  
in any gauge theory with massless gauge bosons (thus QED and QCD in
particular), such scale hierarchies 
give rise to logarithmic enhancements in the matrix elements,
order by order in perturbation theory, which 
ultimately render a truncation of the  perturbative series invalid at
any fixed order.  
This becomes increasingly relevant as the collider CM energy (or other
relevant hard scale) grows, leaving more room for scale hierarchies to
develop beneath it.  

A state-of-the-art collider physics calculation 
includes both a good description of physics at short distances
--- usually represented by perturbative leading-order (LO) or
next-to-leading order (NLO) matrix elements with renormalization-group
(RGE) improved couplings --- 
a good description of the transition from short to long distances,
 which takes proper account of any large scale hierarchies that may
 develop on the way 
 ---- usually represented by leading-log (LL) parton showers incorporating
as many next-to-leading-log (NLL) effects as feasible, as well as
models of other possible perturbatively enhanced aspects such as multiple parton
interactions (MPI) ---  and finally a good 
description of the physics at long distances --- usually represented by 
non-perturbative models of beam remnants, hadronization, and hadron
decays.  

We shall here focus on parton shower algorithms, which provide the
connection between the perturbative fixed-order matrix elements and
the non-perturbative hadronization models, and which thus constitute
an essential ingredient of general-purpose event generators like
\textsc{Herwig}~\cite{Corcella:2000bw,Bahr:2008pv},  
\textsc{Pythia}~\cite{Sjostrand:2006za,Sjostrand:2007gs}, 
or \textsc{Sherpa}~\cite{Gleisberg:2003xi}. 

Parton showers generate 
infinite-order approximations to matrix elements (both real and 
virtual), in such a way as to coincide exactly with the 
matrix elements in the singular limits. The number of singular coefficients
that are reproduced exactly depends on the order of the parton shower; 
thus, an LL parton shower can be expected to generate the
correct coefficients for the ``leading'' matrix-element
singularities (to be defined further below). These terms 
dominate in the limit of infinitely large hierarchy
between the scales of each successive emission, and hence LL showers
are supposed to be exact in this so-called ``strongly ordered'' limit. 
An NLL shower \cite{Kato:1986sg,Kato:1988ii,Kato:1990as,Tanaka:2005rm}
should also generate the correct
coefficients for the next-to-leading singular terms (dominant in 
regions with one less large hierarchy than the strongly ordered
limit), and so on. While the accuracy of the above mentioned general-purpose 
event generators can be debated, we note that none of them systematically include
the tree-level $n\to n+2$ and 1-loop $n\to n+1$ splitting functions,
and hence they are all formally LL. However, also without exception, they
do incorporate a number of non-trivial NLL effects systematically and
usually perform significantly better than corresponding analytical LL
calculations. 
Any proposed algorithm should therefore be subjected to two
basic tests: {\sl 1)} whether it correctly reproduces QCD in the 
LL singular limit, and {\sl 2)} how well it approximates QCD in NLL
singular limits. 

At first order in perturbation theory the matrix elements become
singular in phase space regions corresponding to the emission of
collinear or soft particles. 
The first showering algorithms started from the collinear
factorisation of the matrix elements  
and approximated colour interference effects through angular ordering 
\cite{Marchesini:1983bm,Webber:1983if,Bengtsson:1986hr}. An
alternative approach is the Lund-dipole or dipole-antenna 
shower model, first implemented in
\textsc{Ariadne}~\cite{Gustafson:1986db,Gustafson:1987rq,Andersson:1989ki,Andersson:1988gp,Lonnblad:1992tz},
in which first-order colour interference effects are instead taken into
account by choosing colour-connected \emph{pairs} of
partons to be the radiating entities. 

Here, unfortunately, a digression on nomenclature is necessary. 
These parton pairs were simply called \emph{dipoles} in the original
Lund papers
\cite{Gustafson:1986db,Gustafson:1987rq,Andersson:1989ki,Andersson:1988gp,Lonnblad:1992tz}. 
They are also known as antennae
\cite{Azimov:1984np,Kosower:1997zr,Campbell:1998nn}, 
mostly in fixed-order contexts.
In order to be clear about what we
mean to both communities, we therefore use the term \emph{dipole-antennae}
\cite{Giele:2007di} here for these objects. 
Later, Catani and Seymour invented a related, but different, 
concept which they unfortunately also called dipoles
\cite{Catani:1997vz}. Roughly, a Catani-Seymour dipole corresponds to
one half of a dipole-antenna, partitioned in a way that
isolates the collinear singularities; hence the Catani-Seymour dipole
could also be called a \emph{partitioned-dipole}, as suggested in
\cite{Bern:2008ef}. The problem has been further compounded by a new
generation of shower models based on Catani-Seymour dipoles being
generally referred to as  ``dipole showers''\cite{Nagy:2005aa,Nagy:2006kb,Schumann:2007mg,Dinsdale:2007mf}, 
whereas the same name was
already used by \textsc{Ariadne} to describe its final-state radiation
(FSR) model based on dipole-antennae \cite{Lonnblad:1992tz}. To make the confusion complete, 
while \textsc{Ariadne}'s algorithm for 
final-state radiation is based on a strict dipole-antenna picture, its 
treatment of initial-state radiation (ISR) is somewhat
different. 
We shall nonetheless attempt
to retain some measure of clarity by referring consistently to
the Catani-Seymour type as partitioned-dipoles and to the antenna-type
as dipole-antennae. The two shower types do
exhibit many similarities, so the distinction may mostly
be important to experts, but as with all approximations,  
the devil is in the details.

The last few years have witnessed significant progress in the
improvement of parton shower algorithms.
Based on a proposal by Nagy and Soper\cite{Nagy:2005aa,Nagy:2006kb}
new algorithms based on partitioned-dipoles have been developed
\cite{Schumann:2007mg,Dinsdale:2007mf}. A 
new final-state algorithm relying on the dipole-antenna picture has also been
constructed \cite{Giele:2007di}, with similar properties as the
\textsc{Ariadne} final-state shower, and a complete dipole-antenna
shower model incorporating both ISR and FSR has been developed in 
\cite{Winter:2007ye}.   
These new algorithms have their origin in 
subtraction methods for fixed-order calculations 
\cite{Catani:1997vz,Kosower:1998zr,Phaf:2001gc,Catani:2002hc,GehrmannDeRidder:2005cm,Daleo:2006xa}
and implement in a correct way simultaneously the
soft and the collinear limit in a manner similar to that of 
\textsc{Ariadne}'s FSR model. 
At the same time the new partitioned-dipole algorithms 
(as well as any dipole-antenna algorithm) 
are able to satisfy simultaneously at each step momentum conservation
and the on-shell conditions.
This is possible, because they are based on $2 \rightarrow 3$
splittings, where the spectator can 
absorb the recoil. 
Within the traditional $1 \rightarrow 2$ splitting algorithms it is
impossible to satisfy simultaneously 
momentum conservation and the on-shell conditions for a splitting. 
What has to be done in $1 \rightarrow 2$ splitting algorithms is
either to restore momentum  
conservation by an ad-hoc procedure in the end, as in \textsc{Herwig},
or to restore it more locally during the shower evolution, but still
involving at least a third parton at each step, as in \textsc{Pythia} and
\textsc{Sherpa}. 

In a recent paper Dokshitzer and Marchesini \cite{Dokshitzer:2008ia}
study a soft dipole-type model with recoil effects, which is obtained
from considering multiple antennas in QCD \cite{Bassetto:1984ik}.
The first version of ref.~\cite{Dokshitzer:2008ia} on the archive claimed 
that dipole showers are in conflict with collinear factorization and do
not lead to the correct DGLAP equation.
Ref.~\cite{Dokshitzer:2008ia} motivated us to study this question in detail.
In this paper we show that the above mentioned conclusion is a consequence of an
inappropriate assumption,  
namely the choice of the gluon energy as evolution variable.
We further show that with a proper evolution variable the DGLAP equation is 
reproduced, thus proving that dipole showers, whether of the partitioned
or the antenna type, have the correct LL behaviour, so long as
sensible evolution variables are chosen. Subleading differences
between these shower models will still be present, at the NLL level, which
can first be accessed at $2^\mrm{nd}$ order in perturbative QCD. We therefore
also include a set of explicit comparisons of different shower models to
2$^\mrm{nd}$ order QCD matrix elements. 
We note that in a recent paper Nagy and Soper\cite{Nagy:2009re} have given a strict formal derivation that dipole showers
reduce to the DGLAP equation in the strongly ordered limit.
We further note that the second version of ref.~\cite{Dokshitzer:2008ia} on the archive only claims that models with the
gluon energy as the evolution variable conflict with collinear factorization.
This is in line with the findings of ref.~\cite{Nagy:2009re} and of this paper.

This paper is organised as follows:
In sect.~\ref{sect:basics} we discuss the factorization of tree-level matrix elements in soft and collinear limits.
In sect.~\ref{sect:shower} we review the way a parton shower is obtained from the factorization properties of the matrix elements.
We discuss in detail the choice of the evolution variable and point out that an energy-ordered shower is not compatible
with the collinear limit.
Showers ordered by the transverse momentum or the virtuality are unproblematic.
In sect.~\ref{sect:fragmentation} we consider the evolution of the 
non-singlet quark fragmentation function.
In sect.~\ref{sect:numerical} we compare the analytical result of
the previous section with numerical results obtained from Monte-Carlo
programs based on both partitioned-dipole and dipole-antenna showers.
Again we show that, so long as ``infrared sensible'' 
evolution variables are chosen, 
these showers correctly reproduce the collinear limit. 
Finally, sect.~\ref{sect:conclusions} contains our conclusions.


\section{Basics}
\label{sect:basics}

\begin{figure}[t]
\begin{center}
\includegraphics[bb= 210 580 400 720]{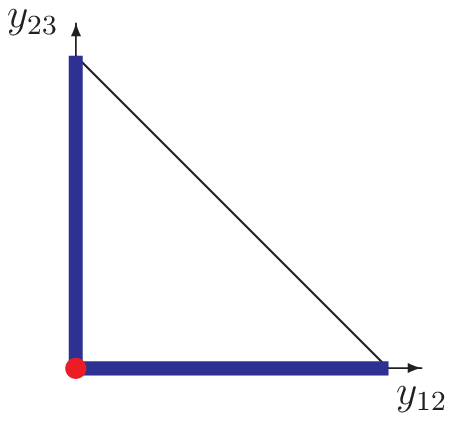}
\end{center}
\caption{The Dalitz plot for the phase space of the unresolved particle.
The location of the soft singularity is shown in red, the location of the collinear singularities in blue.
}
\label{fig_dalitz}
\end{figure}

To set the scene let us consider the matrix element squared for $\gamma^\ast \rightarrow q(p_1) g(p_2) \bar{q}(p_3)$
in four dimensions:
\bq
\left| {\cal A}_3\left(p_1,p_2,p_3\right) \right|^2
 & = & 
 8 e^2 g^2 N_c C_F  
 \left( 
   2 \frac{s_{123}s_{13}}{s_{12}s_{23}}
   + \frac{s_{12}}{s_{23}}
   + \frac{s_{23}}{s_{12}}
 \right).
\eq
In this formula, $e$ denotes the electro-magnetic coupling, $g$ the strong coupling, $N_c=3$ the number of colours and
$C_F=(N_c^2-1)/(2N_c)$.
The invariants are $s_{ij}=(p_i+p_j)^2$ and $s_{ijk}=(p_i+p_j+p_k)^2$.
\\
\\
In the limit where the momentum $p_2$ of the gluon becomes soft, the formula factorises as
\bq
 \lim\limits_{p_2 \rightarrow 0}
\left| {\cal A}_3\left(p_1,p_2,p_3\right) \right|^2
 & = &
 8 \pi \alpha_s \; C_F \;
 \mbox{Eik}(p_1,p_2,p_3) \; 
\left| {\cal A}_2\left(p_1,p_3\right) \right|^2,
\eq
where
\bq
 \mbox{Eik}(p_1,p_2,p_3) & = & 2 \frac{s_{13}}{s_{12}s_{23}},
\eq
and $\alpha_s=g^2/(4\pi)$.
The matrix element squared for
$\gamma^\ast \rightarrow q(p_1) \bar{q}(p_2)$ is given by
\bq
\left| {\cal A}_2\left(p_1,p_2\right) \right|^2
 & = & 
 4 e^2 N_c s_{12}.  
\eq
In the limit where the momentum $p_2$ of the gluon becomes collinear with the momentum $p_1$ of the quark such that
$p_1=z P$ and $p_2=(1-z)P$ we have the
factorization
\bq
 \lim\limits_{p_1 || p_2}
\left| {\cal A}_3\left(p_1,p_2,p_3\right) \right|^2
 & = &
 8 \pi \alpha_s \; C_F \;
 P_{q\rightarrow q g} \;
\left| {\cal A}_2\left(P,p_3\right) \right|^2,
\eq
with the Altarelli-Parisi splitting function
\bq
\label{P_q_qg}
 P_{q\rightarrow q g} & = & 
 \frac{1}{s_{12}} \left( \frac{2}{1-z} - (1+z) \right).
\eq
A similar factorization formula holds for the case where the gluon becomes collinear with the antiquark.
\\
\\
With the help of an antenna function \cite{Kosower:1998zr,Campbell:1998nn} we may combine the three singular limits ($p_2$ soft, $p_2 || p_1$ and $p_2 || p_3$)
into one formula:
\bq
\label{antenna_factorization}
 \lim\limits_{p_2 \;\mbox{\small unresolved}}
\left| {\cal A}_3\left(p_1,p_2,p_3\right) \right|^2
 & = &
 8 \pi \alpha_s \; C_F \;
 A_3^0(p_1,p_2,p_3) \;
\left| {\cal A}_2\left(\tilde{p}_1,\tilde{p}_3\right) \right|^2.
\eq
The antenna function is given by 
\bq
 A_3^0(p_1,p_2,p_3) & = &
 \frac{1}{s_{123}} \left( 2 \frac{s_{13}s_{123}}{s_{12}s_{23}} +
 \frac{s_{12}}{s_{23}} + \frac{s_{23}}{s_{12}} \right), 
\eq
which is exactly the object used in both \textsc{Ariadne} and default \textsc{Vincia}
for $q\bar{q}\to qg\bar{q}$ branchings. 
The momenta $\tilde{p}_1$ and $\tilde{p}_3$ entering the matrix element ${\cal A}_2$
are obtained from $p_1$, $p_2$ and $p_3$ such that they approach the correct limit in all singular limits,
one possibility is \cite{Kosower:1998zr}:
\bq
\tilde{p}_1 & = & \frac{(1+\rho)s_{123} - 2 r s_{23}}{2(s_{123}-s_{23})} p_1 
          + r p_2
          + \frac{(1-\rho)s_{123} - 2 r s_{12}}{2(s_{123}-s_{12})} p_3,
 \nonumber \\
\tilde{p}_3 & = & \frac{(1-\rho)s_{123} - 2 (1-r) s_{23}}{2(s_{123}-s_{23})} p_1 
          + (1-r) p_2
          + \frac{(1+\rho)s_{123} - 2 (1-r) s_{12}}{2(s_{123}-s_{12})} p_3,
\eq
where
\bq
 r= \frac{s_{23}}{s_{12}+s_{23}},
 & &
 \rho = \sqrt{1 + 4 r (1-r) \frac{s_{12}s_{23}}{s_{123} s_{13}} }.
\eq
Note that eq.~(\ref{antenna_factorization}) gives the correct factorization in any singular limit: $p_2$ soft,
$p_2 || p_1$ and $p_2 || p_3$.
Let us introduce the dimensionless quantities
\bq
 y_{12} = \frac{s_{12}}{s_{123}} = 1 - x_3,
 \;\;\;
 y_{23} = \frac{s_{23}}{s_{123}} = 1 - x_1,
 \;\;\;
 y_{13} = \frac{s_{13}}{s_{123}} = 1 - x_2,
\eq
where the $x_i$ are the ordinary energy fractions, evaluated in the CM of
the 3-parton system, 
\begin{equation}
x_i = \frac{2 E_i}{\sqrt{s_{123}}}~.
\end{equation}
Obviously we have $y_{12}+y_{23}+y_{13}=1$ and $x_1+x_2+x_3 = 2$.
The antenna function reads then
\bq
 A_3^0(p_1,p_2,p_3) & = &
 \frac{1}{s_{123}} \left( 2 \frac{y_{13}}{y_{12} y_{23}} + \frac{y_{12}}{y_{23}} + \frac{y_{23}}{y_{12}} \right).
\eq
The unresolved phase space in terms of these variables is
\bq
 d\phi_{unresolved} & = &
 \frac{s_{123}}{16 \pi^2} dy_{12} dy_{23} dy_{13} \delta\left(1-y_{12}+y_{23}+y_{13})\right)
 \Theta(y_{12}) \Theta(y_{23}) \Theta(y_{13}).
\eq
The Dalitz plot for the unresolved phase space is shown in fig~\ref{fig_dalitz}.
The allowed phase space consists of the triangle $y_{12}\ge 0$, $y_{23}\ge 0$, $y_{12}+y_{23}\le 1$.
The soft singularity corresponds to the point $y_{12}=y_{23}=0$, 
while the collinear singularity $p_1||p_2$ corresponds to the line
$y_{12}=0$ (and the collinear singularity $p_2||p_3$ corresponds 
to the line $y_{23}=0$).
\\
\\
We can write the antenna function $A_3^0$ in terms of two dipoles:
\bq
 A_3^0(p_1,p_2,p_3) & = & D_{12,3} + D_{32,1},
\eq
with 
\bq
 D_{12,3} = \frac{1}{s_{123}} \left( \frac{2y_{13}}{y_{12} (y_{12}+y_{32})} + \frac{y_{23}}{y_{12}} \right),
 \;\;\;
 D_{32,1} = \frac{1}{s_{123}} \left( \frac{2y_{13}}{y_{23} (y_{12}+y_{23})} + \frac{y_{12}}{y_{23}} \right).
\eq
In their recent paper Dokshitzer and Marchesini \cite{Dokshitzer:2008ia}
study a model consisting of only the soft part of the dipoles:
\bq
 D^{soft}_{12,3} = \frac{2y_{13}}{y_{12} (y_{12}+y_{23}) s_{123}},
 \;\;\;
 D^{soft}_{32,1} = \frac{2y_{13}}{y_{23} (y_{12}+y_{23}) s_{123}}.
\eq
Dokshitzer and Marchesini actually use slightly different dipoles, obtained upon averaging over the azimuthal
angle of the emitted particle around the emitter.
The differences in the choice for the dipoles will not be relevant to the rest of the paper.
Note that although we label these terms ``soft'', the terms $D^{soft}_{12,3}$ and $D^{soft}_{32,1}$ have a soft and collinear
singularity.
It is also clear, that in the collinear limit $D^{soft}_{12,3}$ does not reduce to $P_{q\rightarrow q g}$, but to
\bq
\label{P_q_qg_soft}
 P^{soft}_{q\rightarrow q g} & = &
 \frac{2z}{(1-z) s_{12}}.
\eq
In statements about shower algorithms reproducing the correct DGLAP equation
in the collinear limit, the term ``correct DGLAP equation'' refers therefore for the model
above to eq.~(\ref{P_q_qg_soft}) and not to eq.~(\ref{P_q_qg}).


\section{Parton shower}
\label{sect:shower}

\begin{figure}[t]
\begin{center}
\includegraphics[bb= 140 580 460 720]{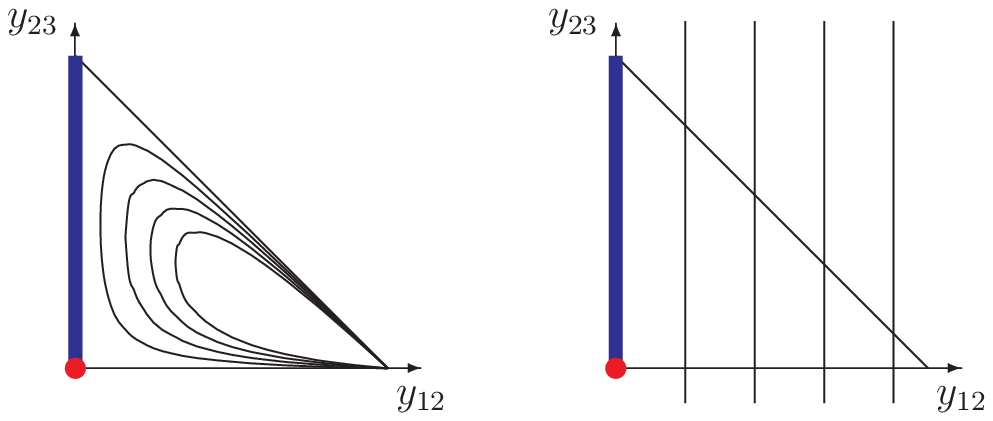}
\end{center}
\caption{Lines of constant shower time $t$ for 
a $k_\perp$-ordered shower (left) and a virtuality-ordered shower (right).
The singular region for the dipole $D_{12,3}$ is shown in blue and red.
The singular region is approached for $t \rightarrow \infty$.
}
\label{fig_constant_time}
\end{figure}

Whereas the factorization formula eq.~(\ref{antenna_factorization}) is exact in all singular limits, 
the r.h.s of eq.~(\ref{antenna_factorization})
does however not necessarily equal the full matrix element on the l.h.s away from the singular limit.
(In the particular example discussed above the full matrix element for
$\gamma^\ast \rightarrow q g \bar{q}$
actually equals the factorized form over the complete
phase space, but this is not the general case.)
Since for high parton multiplicities the full matrix elements are too complicated one approximates in a parton shower
the full matrix elements
by the factorized form over the complete phase space.
Let us stress that this identification is exact in all singular limits and an approximation away from the singular limits.
\\
\\
The antenna function $A_3^0$ and the dipoles $D_{12,3}$ and $D_{32,1}$ are positive definite over the complete
phase space and therefore can be interpreted as a probability distribution for the emission of an additional
particle.
For a parton shower algorithm
we introduce two variables $t$ (``shower time'') and $z$ (``momentum fraction'').
The shower time $t$ gives the scale at which the next splitting occurs, the variable $z$ describes for a splitting
$a \rightarrow b c$ the momentum fraction of the daughter $b$ with respect to the mother $a$.
For a complete description of a splitting we need in principle a third variable $\phi$, but this variable will
not be relevant for the discussion of this paper and we suppress it. Therefore we can restrict ourselves
to a two-dimensional space parametrized by $(t,z)$ or $(y_{12},y_{23})$ as in fig.~\ref{fig_dalitz}.
The choice for $t$ (and $z$) is not unique.
If we focus on the dipole $D_{12,3}$ with singularities for $s_{12} \rightarrow 0$
possible choices are
\bq
\label{kt_ordered}
 t_{\tilde{1}\tilde{3}} & = & - \ln \frac{-k_\perp^2}{Q^2}
 =
 - \ln \left( \frac{y_{12} y_{23} y_{13}}{(1-y_{12})^2} \frac{s_{\tilde{1}\tilde{3}}}{Q^2} \right)
\eq
for a $k_\perp$-ordered shower
or
\bq
\label{virtuality_ordered}
 t_{\tilde{1}\tilde{3}} & = & - \ln \frac{s_{12}}{Q^2}
 =
 - \ln \left( y_{12} \frac{s_{\tilde{1}\tilde{3}}}{Q^2} \right)
\eq
for a virtuality-ordered shower.
The quantity $Q^2$ in eq.~(\ref{kt_ordered}) and
 eq.~(\ref{virtuality_ordered}) is a fixed reference scale, usually
 taken to be the centre-of-mass energy squared of the showering system.
The shower time $t$ takes values between a starting time $t_0$ and $+\infty$.
As a larger values of $t$ corresponds to lower scales we have to require that the singular region is
contained in the region defined by $t \rightarrow \infty$.
In fig.~\ref{fig_constant_time} we show in the Dalitz plot lines of constant $t$ for the definitions
as in eq.~(\ref{kt_ordered}) and eq.~(\ref{virtuality_ordered}).
It is clear that the collinear singular region $y_{12}=0$ is contained
for both definitions in the region defined by $t\rightarrow \infty$. 

\begin{figure}[t]
\begin{center}
\includegraphics[bb= 140 580 460 720]{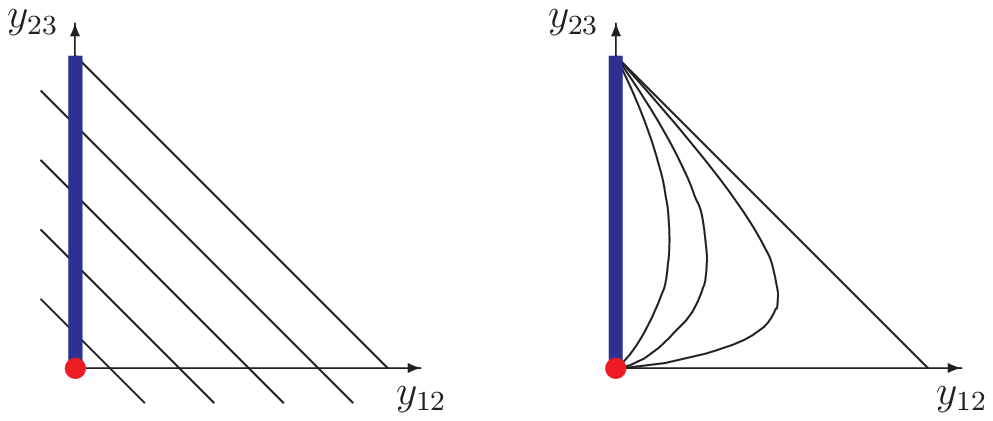}
\end{center}
\caption{Lines of constant shower time $t$ for 
an energy-ordered shower (left) and an angular-ordered shower (right).
The singular region for the dipole $D_{12,3}$ is shown in blue and red.
For the energy-ordered shower lines of constant $t$ intersect the singular region for finite $t$.
For the angular-ordered shower the two singular points $(y_{12},y_{23})=(0,0)$ and $(y_{12},y_{23})=(0,1)$ are reached for finite shower time.
}
\label{fig_constant_time_energy_ordered}
\end{figure}

Associated to the scale $t$ is the Sudakov factor $\Delta_{12,3}$, giving the probability that no emission occurs between
the scales $t_0$ and $t$
\bq
\label{sudakov}
 \Delta_{12,3}(t_0,t) & = &
 \exp\left( - \int\limits_{t_0}^{t} dt'  
              \int d\phi_{unres} \delta\left(t'-t_{\tilde{1}\tilde{3}} \right) 
              8 \pi \alpha_s C_F D_{12,3}\right),
\eq
with $t_{\tilde{1}\tilde{3}}$ given by
eq.~(\ref{kt_ordered}) or eq.~(\ref{virtuality_ordered}).
The derivative of the Sudakov factor with respect to $t$ gives the probability of a splitting at the shower time
$t$:
\bq
- \frac{d}{dt} \Delta_{12,3}(t_0,t) & = &
 \frac{\alpha_s}{2\pi} C_F \int d\phi_{unres} \delta\left(t-t_{\tilde{1}\tilde{3}} \right)
 s_{123} D_{12,3}
\eq
Working this out for a virtuality-ordered shower and setting
 $z=1-y_{23}$ one obtains 
\bq
- \frac{d}{dt} \Delta_{12,3}(t_0,t) & = &
 \frac{\alpha_s}{2\pi} C_F 
 \int\limits_{\delta}^1 dz \left( \frac{2}{1-z+\delta} - 1 - z \right),
 \;\;\;
 \delta = \frac{Q^2}{s_{\tilde{1}\tilde{3}}} e^{-t}.
\eq
At finite shower time $t$ the splitting probability is finite, as it should be.
In the limit $t\rightarrow \infty$ one recovers the DGLAP equation:
\bq
 \lim\limits_{t\rightarrow \infty} \left( - \frac{d}{dt} \Delta_{12,3}(t_0,t) \right)
 & = & 
 \frac{\alpha_s}{2\pi} C_F 
 \int\limits_0^1 dz \left( \frac{2}{1-z} - 1 - z \right).
\eq
A similar analysis can be carried out for a $k_\perp$-ordered shower.
\\
\\
In their paper Dokshitzer and Marchesini \cite{Dokshitzer:2008ia} did not use 
the transverse momentum $k_\perp$ or the virtuality as evolution variable.
Instead they chose energy of the emitted gluon (in the rest frame of the dipole)
as evolution variable.
In our notation this amounts to the choice
\bq
\label{energy_ordered}
 t_{\tilde{1}\tilde{3}} & = & - \ln \frac{E_g^2}{Q^2}
 =
 - \ln \left( \frac{(y_{12}+y_{23})^2}{4} \frac{s_{\tilde{1}\tilde{3}}}{Q^2} \right).
\eq
For this choice the lines of constant shower time $t$ are shown in fig.~\ref{fig_constant_time_energy_ordered}.
In this case $t \rightarrow \infty$ corresponds to the single point
$y_{12}=y_{23}=0$ in phase space and lines of constant $t$ intersect the collinear region
for finite $t$.
We can therefore not expect to obtain a shower algorithm which is correct in the collinear limit
based on this evolution variable.
Indeed, a short calculation shows in this case
\bq
\label{eq_energy_ordered}
- \frac{d}{dt} \Delta_{12,3}(t_0,t) & = &
 \frac{\alpha_s}{2\pi}  C_F  
 \theta(1-y_{max})
 \int\limits_0^{y_{max}}  
 \frac{dy_{23}}{y_{max}-y_{23}} \left[ 1 - y_{max} \left( 1 - \frac{1}{2} y_{23} \right) \right],
 \nonumber \\
 y_{max} & = & 2 \sqrt{\frac{Q^2}{s_{\tilde{1}\tilde{3}}} } e^{-\frac{t}{2}}.
\eq
There are several problems related to an energy-ordered shower:
For finite shower time $t$ the splitting probability is infinite, due to the $1/(y_{max}-y_{23})$
singularity in the integrand. In the limit $t\rightarrow \infty$ the integration over $y_{23}$ reduces
to a point and not to the integral over the splitting function.
The integrand of eq.~(\ref{eq_energy_ordered}) bears no resemblance to the Altarelli-Parisi splitting function.
These deficiencies are all related to the inappropriate choice of the shower evolution variable.
\\
\\
For completeness we also show the corresponding plot for an angular-ordered shower in fig.~\ref{fig_constant_time_energy_ordered}.
Lines of constant angle are given by
\bq
 1-\cos\theta_{12} & = & 2 \frac{y_{12}}{(y_{12}+y_{23})(1-y_{23})}
\eq
and the corresponding definition of the shower time $t$ is given by
\bq
 t_{\tilde{1}\tilde{3}} & = & 
 - \ln \left( \frac{y_{12}}{(y_{12}+y_{23})(1-y_{23})} \frac{s_{\tilde{1}\tilde{3}}}{Q^2} \right).
\eq
The points $(y_{12},y_{23})=(0,0)$ and $(y_{12},y_{23})=(0,1)$ are also reached for finite
shower time $t$ and angular-ordered showers have to introduce a cut-off on the variable $z$ to
avoid these points.
In more detail we have for an angular-ordered shower
\bq
\label{eq_angular_ordered}
- \frac{d}{dt} \Delta_{12,3}(t_0,t) & = &
 \frac{\alpha_s}{2\pi}  C_F  
 \theta(1-y_{max})
 \int\limits_0^1 dy_{23}
 \left[ 2 \frac{1-y_{23}}{y_{23}} \frac{1-y_{max}}{1-(1-y_{23})y_{max}} + \frac{y_{23}}{1-(1-y_{23})y_{max}} \right],
 \nonumber \\
 y_{max} & = & \frac{Q^2}{s_{\tilde{1}\tilde{3}}} e^{-t}.
\eq
As already mentioned above, we have for an angular ordered shower for finite $t$ 
an infinite splitting probability due to the soft singularity at $y_{23}=0$.
In angular-ordered shower programs this situation is usually handled by introducing an
ad-hoc cut-off on the variable $y_{23}$.
In the limit $t\rightarrow \infty$ and setting $z=1-y_{23}$ we recover the DGLAP equation:
\bq
\lim\limits_{t\rightarrow\infty} \left( - \frac{d}{dt} \Delta_{12,3}(t_0,t) \right) & = &
 \frac{\alpha_s}{2\pi}  C_F  
 \int\limits_0^1 dz
 \left( \frac{2}{1-z} -1 -z \right).
\eq


\section{The non-singlet quark fragmentation function}
\label{sect:fragmentation}

In this section we review the relevant formulae for the evolution of the non-singlet quark
fragmentation function.
The DGLAP evolution equation for the non-singlet quark fragmentation function $d(x,Q^2)$
is
\bq
 Q^2 \frac{d}{dQ^2} d(x,Q^2)
  & = &
 - \int\limits_x^1 \frac{dz}{z} \frac{\alpha_s}{2\pi} C_F P_{qq}(z) d\left(\frac{x}{z},Q^2\right),
\eq
where $P_{qq}$ is the regularized splitting function
\bq
 P_{qq}(z) & = & 
 \left. \frac{2}{1-z} \right|_+ + \frac{3}{2} \delta(1-z) - \left(1+z\right).
\eq
To a function $f(x)$ we denote the Mellin transform by
\bq
 \tilde{f}(N) & = & \int\limits_0^1 dx \; x^{N-1} f(x).
\eq
In Mellin space the evolution equation for the non-singlet quark fragmentation function factorises:
\bq
 Q^2 \frac{d}{dQ^2} \tilde{d}(N,Q^2)
  & = &
 - \frac{\alpha_s}{2\pi} C_F \tilde{P}_{qq}(N) \tilde{d}(N,Q^2),
\eq
with
\bq
 \tilde{P}_{qq}(N) & = & -2 S_1(N-1) + \frac{3}{2} - \frac{1}{N} - \frac{1}{N+1}.
\eq
$S_1(N-1)$ is the harmonic sum
\bq
 S_1(N-1) & = & \sum\limits_{j=1}^{N-1} \frac{1}{j}.
\eq
For large $N$, $S_1(N-1)$ diverges logarithmically.
In Mellin space the evolution equation can be solved analytically
\bq
\tilde{d}(N,Q^2) & = & 
 \left( 1 + \frac{\alpha_s(Q_0)}{4\pi} \beta_0 \ln \frac{Q^2}{Q_0^2} \right)^{-\frac{2}{\beta_0} C_F \tilde{P}_{qq}(N)} \tilde{d}(N,Q_0^2),
\eq
where $\beta_0=11/3 C_A -4/3 T_r N_f$ is the first coefficient of the beta-function.
We are also interested in a toy model with $\alpha_s=const$, in this case the solution is given by
\bq
\label{solution_const_alpha_s}
\tilde{d}(N,Q^2) & = & 
 \left( \frac{Q^2}{Q_0^2} \right)^{-\frac{\alpha_s}{4\pi} 2 C_F \tilde{P}_{qq}(N)} \tilde{d}(N,Q_0^2).
\eq
The initial condition 
\bq
\label{initial_condition}
 d(x,Q_0^2) & = & \delta(1-x)
\eq
corresponds in Mellin space to
\bq
 \tilde{d}(N,Q_0^2) & = & 1.
\eq
Finally we are interested in the $x$-space result for the quark fragmentation function $d(x,Q^2)$
for values of $x$ close to $1$ and with $\alpha_s=const$.
In this region only soft gluons have been emitted.
With the ansatz \cite{Altarelli:1981ax}
\bq
 d(x,Q^2) & = & A(Q^2) \left(1-x\right)^{B(Q^2)}
\eq
and neglecting terms which vanish in the limit $x\rightarrow 1$
one finds the equation
\bq
 \frac{1}{A(Q^2)} \frac{dA(Q^2)}{d\ln Q^2} 
 + \frac{\alpha_s C_F}{\pi} \left[ \frac{3}{4} -\gamma_E - \psi\left(B(Q^2)+1\right) \right]
 + \ln(1-x) \left[ \frac{dB(Q^2)}{d\ln Q^2} + \frac{\alpha_s C_F}{\pi} \right] & = & 0.
\;\;\;\;\;\;
\eq
The coefficient of $\ln(1-x)$ and the term independent of $\ln(1-x)$ have to vanish independently.
With the initial condition eq.~(\ref{initial_condition}) one then obtains
\bq
\label{res_x_space}
\lefteqn{
 \ln d(x,Q^2) = } & &  
\nonumber \\
 & &
 - \left[ 1 + \frac{\alpha_s C_F}{\pi} \ln \frac{Q^2}{Q_0^2} \right] \ln(1-x)
 - \frac{\alpha_s C_F}{\pi} \left( \frac{3}{4} - \gamma_E \right) \ln \frac{Q^2}{Q_0^2}
 - \ln \Gamma\left( - \frac{\alpha_s C_F}{\pi} \ln
 \frac{Q^2}{Q_0^2} \right). \label{eq:dAna}
\eq
This solution is valid for
\bq
\label{validity}
 (1-x) \ll 1 \;\;\;\mbox{and}\;\;\;
 \frac{\alpha_s}{\pi} \ln \frac{1}{1-x} \ll 1.
\eq


\section{Numerical studies}
\label{sect:numerical}

\subsection{The quark fragmentation function}

\begin{figure}[t]
\begin{center}
\includegraphics[bb= 125 460 490 710,width=0.45\textwidth]{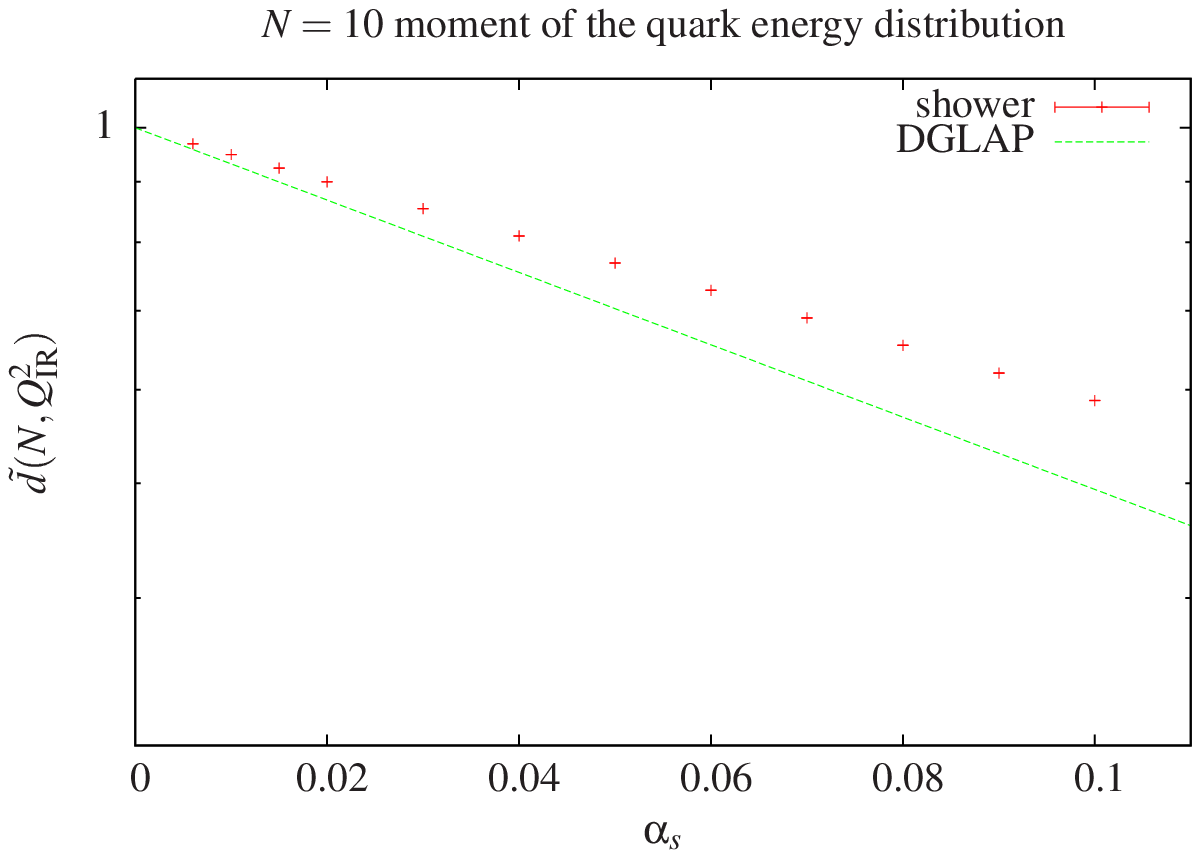}
\includegraphics[bb= 125 460 490 710,width=0.45\textwidth]{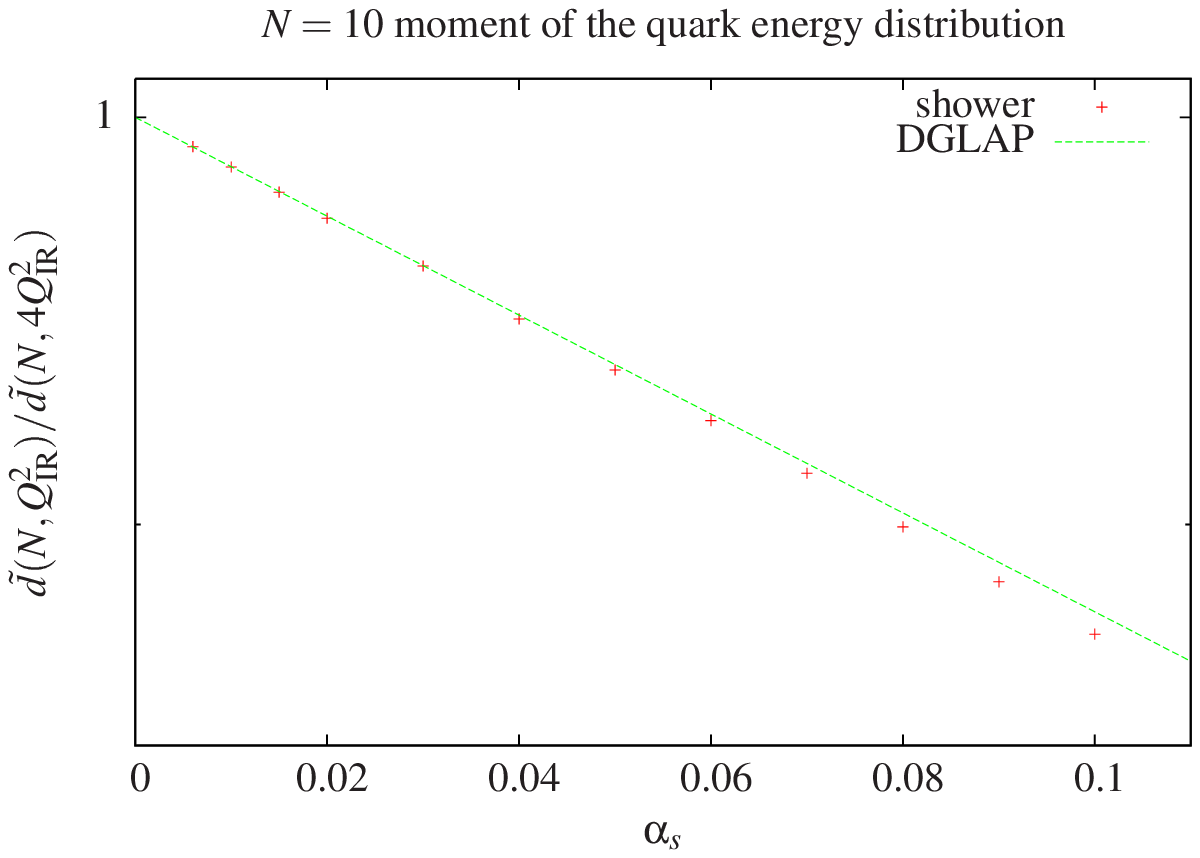}
\end{center}
\caption{
The $10$-th moment of the quark energy distribution for various values of $\alpha_s$,
obtained from a process with centre-of-mass energy $Q=m_Z$.
The left figure shows the moment at $Q_{\mathrm{IR}}=1\;\mbox{GeV}$, the right figure
shows the ratio between the values at $Q_{\mathrm{IR}}=1\;\mbox{GeV}$ and $Q_{\mathrm{IR}}=2\;\mbox{GeV}$.
}
\label{fig_N10}
\end{figure}

In this section we study the quark energy distribution 
in Monte Carlo events obtained from a shower simulation, starting from the
hard matrix element $e^+ e^- \rightarrow q \bar{q}$.
As centre-of-mass energy we take $Q=m_Z$, unless indicated otherwise. The $(N-1)$-th moment of the
quark energy distribution at the scale $Q_j$
is just
\bq
 \tilde{d}(N,Q_j^2).
\eq
Our main interest is the comparison between the numerical shower program and the analytical
result from the DGLAP equation. For this comparison it is sufficient to consider a toy model
with $\alpha_s=const$. We set $\alpha_s=0.1$.
For the numerical result we first generate quark-antiquark events according to the hard matrix element,
then start the shower at a scale $Q_0$ and run the shower to the lower scale $Q_{\mathrm{IR}}$.
For a $k_\perp$-ordered shower starting from a process at the centre-of-mass energy $Q$
the upper limit on $Q_0$ is given by $Q_{0,max} = Q/2$.
We then calculate the energy fraction of the quark (additional quarks obtained from 
$g \rightarrow q \bar{q}$ splittings are not relevant to the discussion here):
\bq
 x & = & \frac{2E_q}{Q}.
\eq
This defines a distribution in $x$. 
In addition, we can simultaneously bin the moments of this distribution.
\\
\\
We perform several comparisons between the numerical shower program and the analytical
result from the DGLAP equation.
In Mellin space the DGLAP equation is an ordinary differential equation and the numerical shower program
has to reproduce this equation in the strongly ordered limit.
Strongly-ordered implies that the scale of successive emissions satisfy
\bq
\label{eq_strongly_ordered}
 ... \gg Q^2_{j-1} \gg Q^2_j \gg Q^2_{j+1} \gg ...
\eq
We can ensure these conditions by starting the shower at a rather low scale $Q_0=2\;\mbox{GeV}$
and run the shower only for a short interval to $Q_{\mathrm{IR}}=1\;\mbox{GeV}$.
The low starting scale $Q_0$ ensures $Q^2 \gg Q_0^2$ and no emissions with a scale larger than
$Q_0$ are generated.
The short interval ensures that the number of events with two or more emissions is negligible and
almost all events will have either zero or one shower emission.
This is necessary since although in a shower successive emissions are ordered
\bq
\label{eq_ordered}
 ... > Q^2_{j-1} > Q^2_j > Q^2_{j+1} > ...,
\eq
condition~(\ref{eq_ordered}) does not exclude successive emissions to be of the same order
$Q^2_j = {\cal O}(Q^2_{j-1})$.
Running the shower over the short interval from $2\;\mbox{GeV}$ to $1\;\mbox{GeV}$ 
tests therefore if the emission of a single particle correctly approaches the DGLAP limit.
\begin{figure}[t]
\begin{center}
\includegraphics[bb= 125 460 490 710,width=0.9\textwidth]{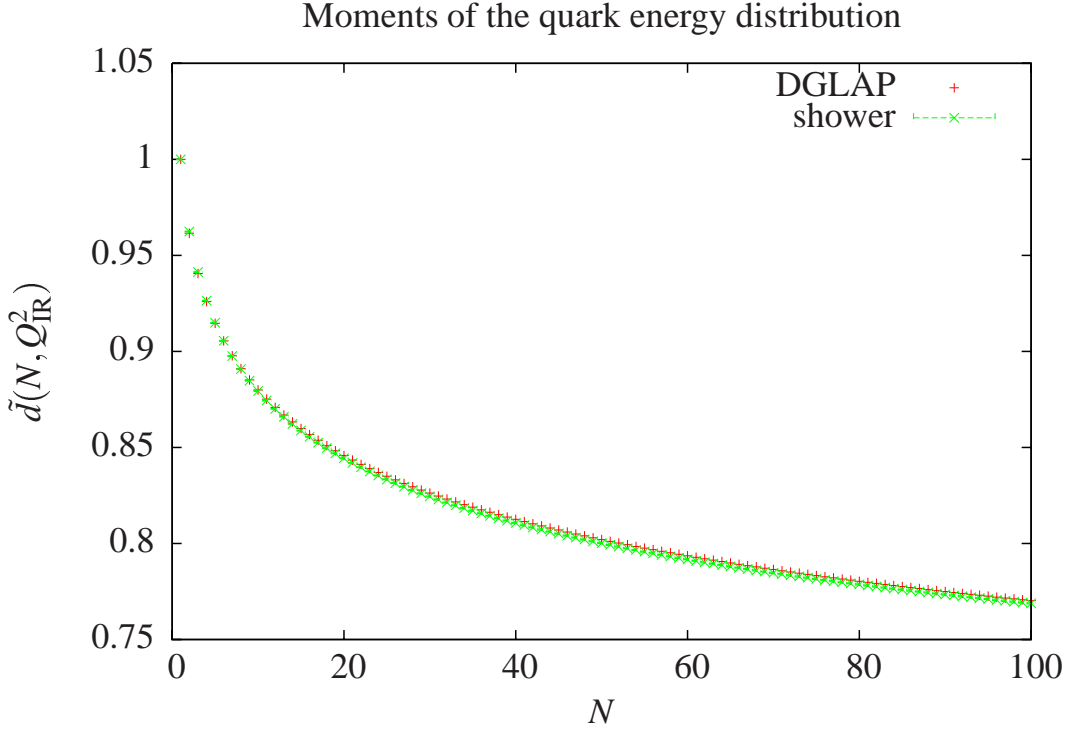}
\end{center}
\caption{
Moments of the quark energy distribution at $Q_{\mathrm{IR}}=1\;\mbox{GeV}$ obtained from 
starting the evolution at $Q_0=2\;\mbox{GeV}$.
}
\label{fig_2GeV}
\end{figure}
Fig.~\ref{fig_2GeV} shows for this case the comparison between the numerical shower program and the
analytical solution eq.~(\ref{solution_const_alpha_s}).
We observe an excellent agreement.
\\
\\
We then compare the numerical shower program and the analytical
result from the DGLAP equation for a larger interval for the evolution.
We start the shower at the hard scale $Q_0=Q_{0,max}$ and run to the low scale $Q_{\mathrm{IR}}=1\;\mbox{GeV}$.
We compare again the moments of the quark energy distribution.
We do this for the centre-of-mass energies $Q=m_Z$, $Q=1\;\mbox{TeV}$, $Q=10\;\mbox{TeV}$
and $Q=100\;\mbox{TeV}$.
We do not expect perfect agreement, since now the shower may generate emissions with a scale smaller but
comparable to the previous one.
However both the shower and the DGLAP equation resum the leading logarithm. In the limit where this logarithm
is large against other terms, the results should agree.
\begin{figure}[t]
\begin{center}
\includegraphics[bb= 125 460 490 710,width=0.45\textwidth]{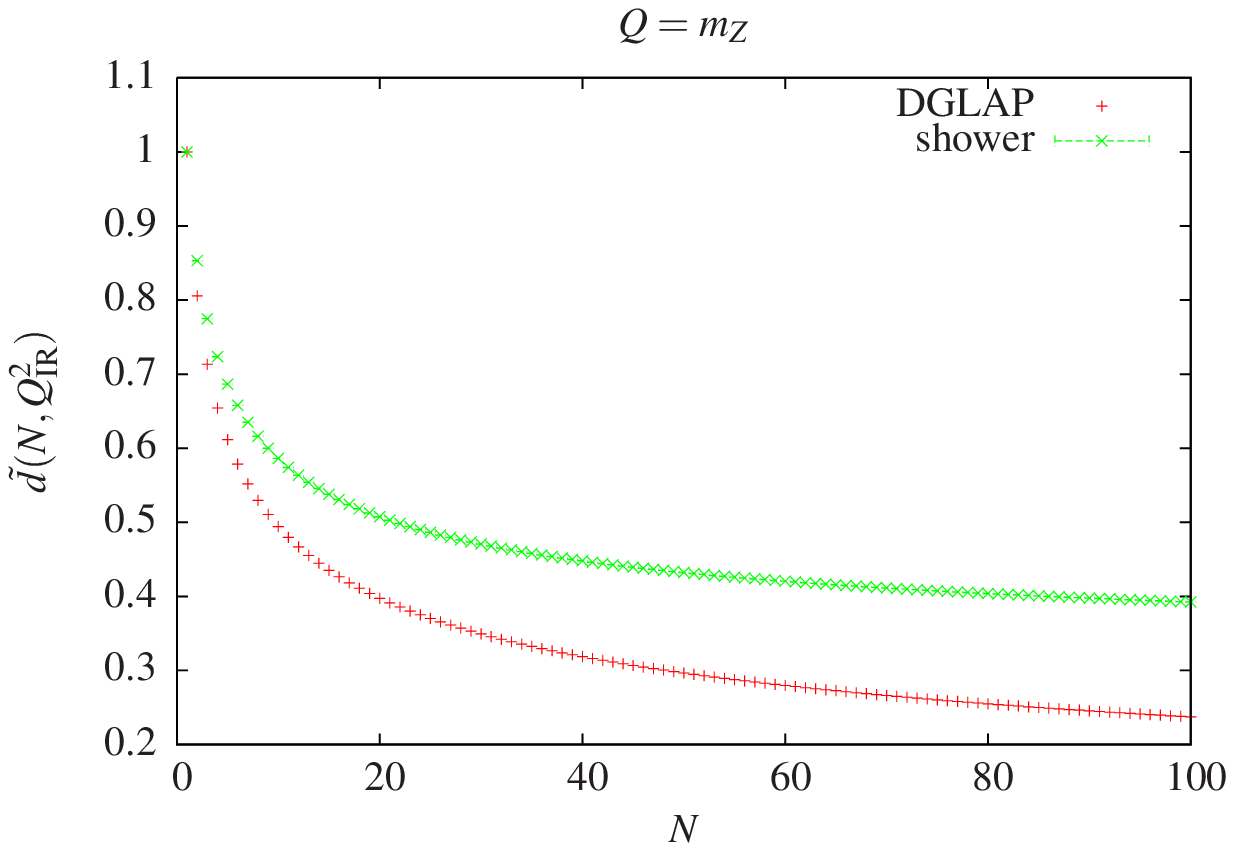}
\includegraphics[bb= 125 460 490 710,width=0.45\textwidth]{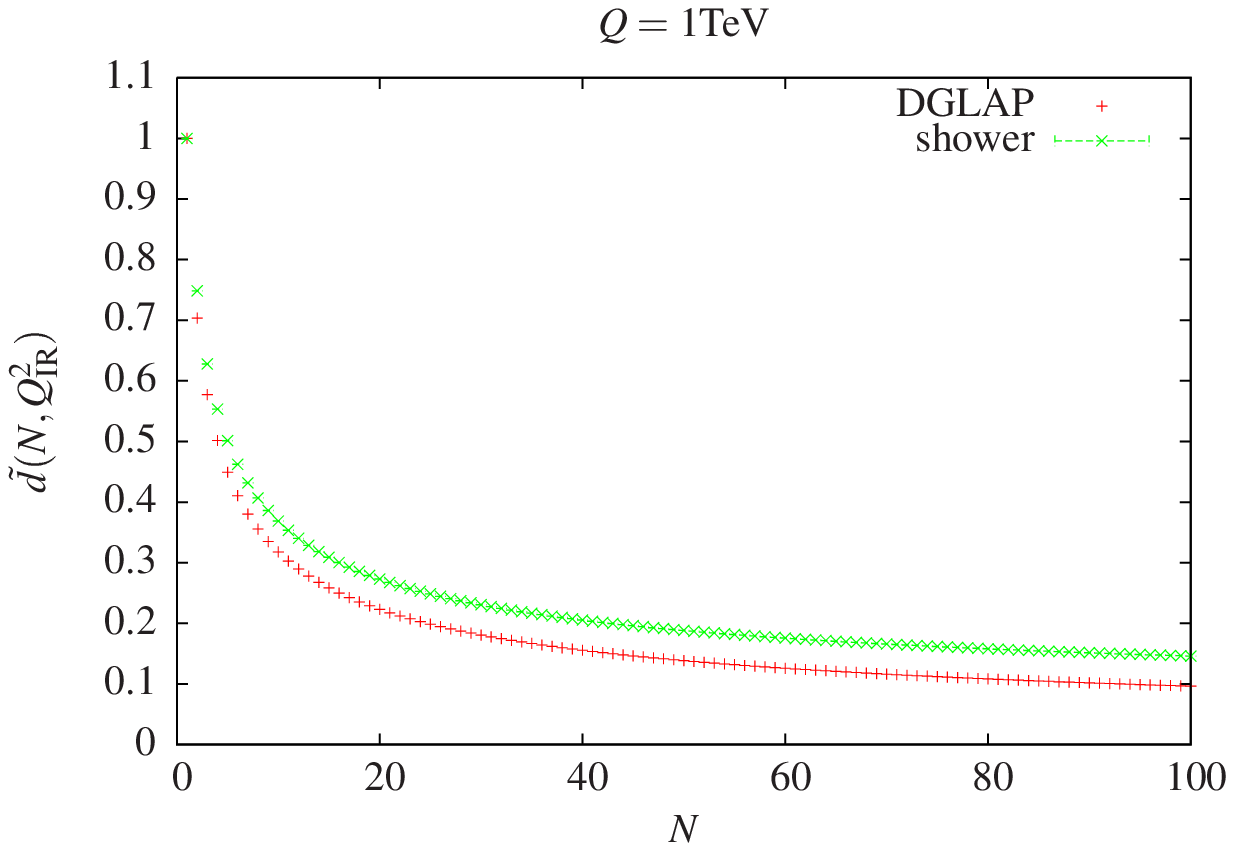}
\\
\includegraphics[bb= 125 460 490 710,width=0.45\textwidth]{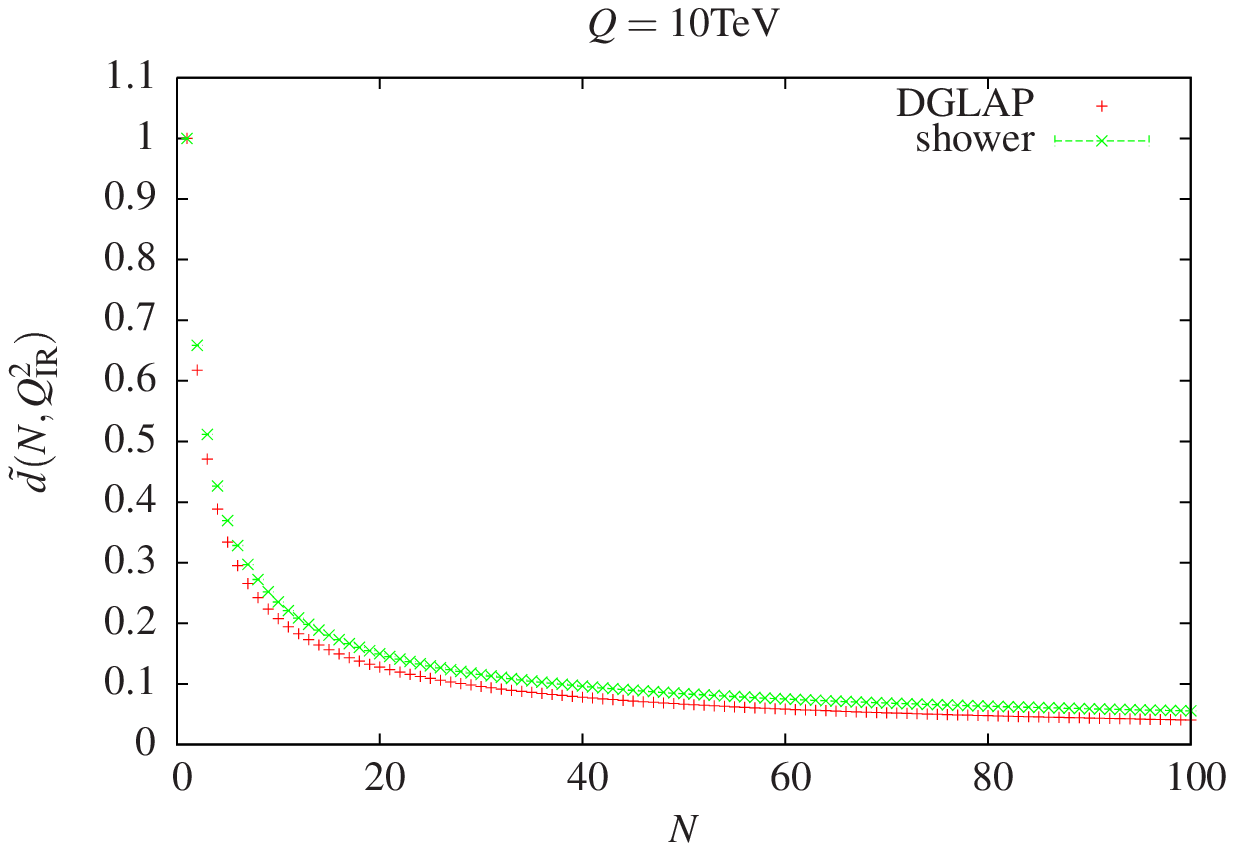}
\includegraphics[bb= 125 460 490 710,width=0.45\textwidth]{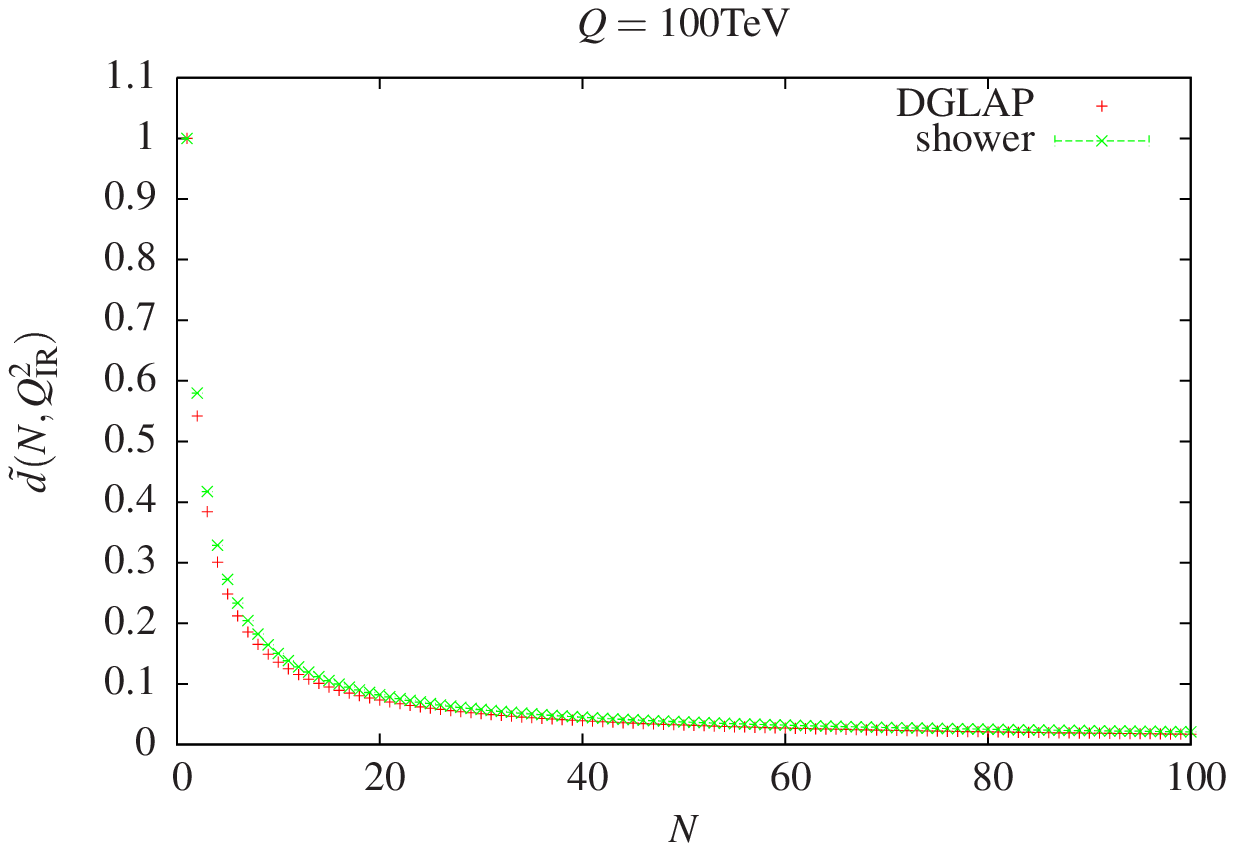}
\end{center}
\caption{
Moments of the quark energy distribution at $Q_{\mathrm{IR}}=1\;\mbox{GeV}$ for various
centre-of-mass energies: $Q=m_Z$, $Q=1\mbox{TeV}$, $Q=10\mbox{TeV}$ and $Q=100\mbox{TeV}$.
}
\label{fig_moments_long_evolv}
\end{figure}
Fig.~\ref{fig_moments_long_evolv} shows the comparison between the numerical shower program and the
analytical solution eq.~(\ref{solution_const_alpha_s}) 
for the centre-of-mass energies  $Q=m_Z$, $Q=1\;\mbox{TeV}$, $Q=10\;\mbox{TeV}$ and $Q=100\;\mbox{TeV}$.
The starting scale of the shower is always $Q_0=Q_{0,max}=Q/2$. For the final scale of the shower
the value $Q_{\mathrm{IR}}=1\;\mbox{GeV}$ is always used.
We observe that for large values of
\bq
 \ln \frac{Q_0^2}{Q_{\mathrm{IR}}^2}
\eq
the two results approach each other.
\\
\\
Fig.~\ref{fig_N10} shows the $10$-th moment of the quark energy distribution as a function of
the (fixed) value of $\alpha_s$.
The centre-of-mass energy is $Q=m_Z$ and the starting scale of the shower $Q_0=Q_{0,max}=Q/2$.
As the low scale $Q_{\mathrm{IR}}=1\;\mbox{GeV}$ is used.
From eq.~(\ref{solution_const_alpha_s}) we expect on a logarithmic scale a linear relationship
with respect to the variation of $\alpha_s$:
\bq
\label{slope}
\ln \tilde{d}(N,Q_{\mathrm{IR}}^2) & = & 
 -\frac{\alpha_s}{4\pi} 2 C_F \tilde{P}_{qq}(N) \ln \frac{Q_{\mathrm{IR}}^2}{Q_0^2}.
\eq
We observe in the left plot of fig.~\ref{fig_N10} that both the numerical result from the shower program
and the theoretical curve give straight lines. However the slope is slightly different.
From eq.~(\ref{slope}) we see that the slope depends on the value of the hard scale $Q_0$.
To eliminate the dependence on $Q_0$ we show in the right plot of fig.~\ref{fig_N10} the ratio
$\tilde{d}(N,Q_{\mathrm{IR}}^2) / \tilde{d}(N,4 Q_{\mathrm{IR}}^2)$ of the $10$-th moment at the low scales
$Q_{\mathrm{IR}}$ and $2 Q_{\mathrm{IR}}$. In this ratio the dependence on $Q_0$ drops out:
\bq
\ln \frac{\tilde{d}(N,Q_{\mathrm{IR}}^2)}{\tilde{d}(N,4 Q_{\mathrm{IR}}^2)} & = & 
 \frac{\alpha_s}{\pi} C_F \tilde{P}_{qq}(N) \ln 2.
\eq
We observe an excellent agreement.
\\
\\
As a further comparison we now study the quark energy distribution in $x$-space for values of $x$ close to $1$.
This region is sensitive to the emission of soft gluons.
We start the shower at the hard scale $Q_0=Q_{0,max}=Q/2=m_Z/2$, and use as the final scale of the shower 
$Q_{\mathrm{IR}}=1\;\mbox{GeV}$.
\begin{figure}[t]
\begin{center}
\includegraphics[bb= 125 460 490 710,width=0.9\textwidth]{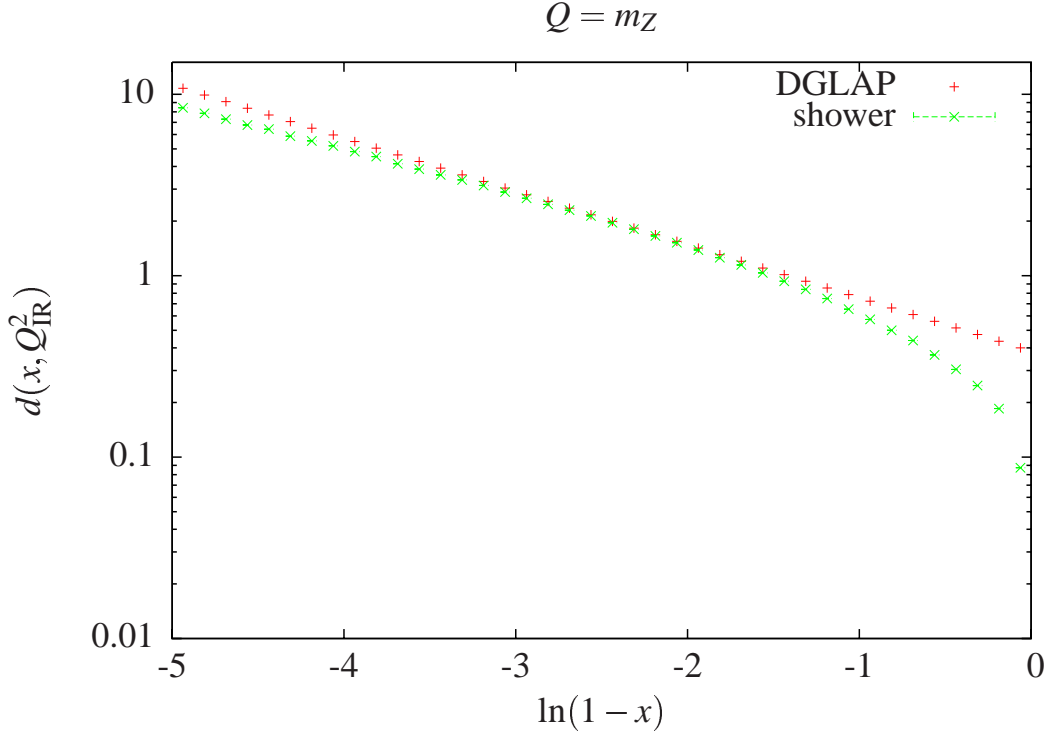}
\end{center}
\caption{
The quark energy distribution for small values of $(1-x)$ at $Q_{\mathrm{IR}}=1\;\mbox{GeV}$ 
for the centre-of-mass energy $Q=m_Z$. 
}
\label{fig_x_long_evolv}
\end{figure}
Fig.~\ref{fig_x_long_evolv} shows the comparison between the numerical shower program and the
analytical solution eq.~(\ref{res_x_space}).
We observe a good agreement.
We would like to make a comment: The analytical solution gives a linear relation
\bq
 \ln d(x,Q^2) & = & A \ln(1-x) + B,
\eq
shown as a straight line in fig.~\ref{fig_x_long_evolv}. 
The validity of the analytical solution according to eq.~(\ref{validity}) is 
restricted to the region
\bq
  -\frac{\pi}{\alpha_s} \ll \ln(1-x) \ll 0~.
\eq
Similar, the shower does not generate emissions below a certain value of $\ln(1-x_{max})$.
Values of $\ln(1-x)$ below that value would correspond to emissions with a scale less than
$Q_{\mathrm{IR}}$.
As a consequence there are for any finite value of $Q_{\mathrm{IR}}$ events, which didn't radiate at all.
(The fraction of these events is determined by the Sudakov factor at the scale $Q_{\mathrm{IR}}$.)
In fig.~\ref{fig_x_long_evolv} we have normalized the shower result to the number of events which emitted
at least one additional parton.

To further test the validity of our conclusions under systematic variations
of the shower assumptions, we use the capabilities of the 
\textsc{Vincia} plug-in to the \textsc{Pythia 8} 
generator~\cite{Giele:2007di,Sjostrand:2007gs}, which
offers the possibility to use arbitrary radiation functions in
conjunction with several different evolution variables within a
dipole-antenna shower context. We also compare to the standard
\textsc{Pythia 8} $\pT{}$-ordered shower \cite{Sjostrand:2004ef}, which
represents a hybrid between the parton and dipole approaches. 
In both programs, we switch off $g\to q\bar{q}$
branchings, use a fixed $\alpha_s=0.1$ 
and take the starting scale to be $Q=10^3$ GeV as default, with a
hadronization scale (infrared cutoff) of $Q_{\mathrm{IR}} = 1$~GeV. In
\textsc{Vincia}, we further switch off matching to matrix elements and
matching to $2\to4$ antenna functions.  

\begin{figure}[t]
\begin{center}
\includegraphics[bb= 140 580 460 720]{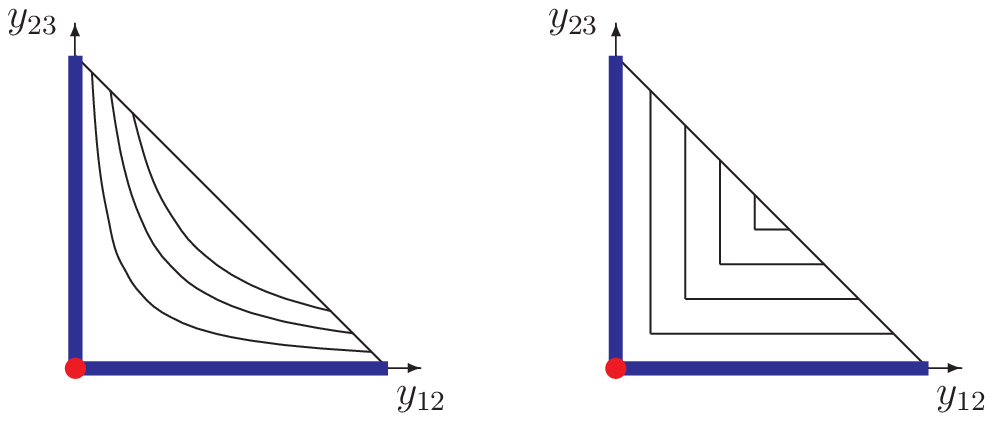}
\end{center}
\caption{Lines of constant shower time $t$ for 
a $p_\perp$-ordered shower (left) and an antenna-mass-ordered shower (right).
The singular region for the antenna $A(y_{12},y_{23})$ is shown in blue and red.
The singular region is approached for $t \rightarrow \infty$.
}
\label{fig:evant}
\end{figure}

In \textsc{Vincia}, we consider two different dipole-antenna 
evolution variables: transverse momentum
(identical to the \textsc{Ariadne} evolution variable \cite{Lonnblad:1992tz}) 
and antenna-mass, referred to as Type I and Type II evolution,
respectively. In terms of colour-ordered triplets of parton momenta,
these variables are defined as follows:
\begin{eqnarray}
Q_{\mathrm{I}}^2(p_1,p_2,p_3) = 4 \frac{s_{12}s_{23}}{s_{123}} \equiv 4 \pT{}^2~,\label{eq:type1}\\
Q_{\mathrm{II}}^2(p_1,p_2,p_3) = 2 \min(s_{12},s_{23}) \equiv 2 m_{\mathrm{ant}}^2,\label{eq:type2} 
\end{eqnarray}
where the normalizations are chosen such that the maximal value of
the evolution variable is $s_{123}$ in both cases. Contours for
constant values of these variables are shown in fig.~\ref{fig:evant}.

The most general form for a leading-log antenna
function (dipole-antenna splitting function) 
for massless parton splitting is represented by a double Laurent
series in the two branching invariants \cite{Bern:2008ef},  
\begin{equation}
A(y_{12},y_{23} ; s_{123}) = \ \ \ \frac{4\pi\alpha_s {\cal C}}{s_{123}}
\!\!\sum_{\alpha,\beta=-1}^\infty\!\! C_{\alpha,\beta} \ 
y_{12}^\alpha\ y_{23}^{\beta} ~~~~~,~~~
\mbox{with}~~~y_{ij} = \frac{s_{ij}}{s_{123}} \le 1 ~~~,
\label{eq:a}
\end{equation}
where ${\cal C}$ is the colour factor. 
We here consider 3 different choices for these functions, the
Gehrmann-de-Ridder-Glover ones (``GGG'')
\cite{GehrmannDeRidder:2005cm}, which are the defaults in
\textsc{Vincia}, and systematically high and low variations, MIN and
MAX, respectively, with coefficients for the $q\qbar\to qg\qbar$ and
$qg\to qgg$ functions given in Tab.~\ref{tab:coefficients}. We note
that the default GGG $q\qbar\to qg\qbar$ function is identical to the
corresponding function in \textsc{Ariadne} and reproduces the $Z\to
qg\qbar$ tree-level matrix element exactly. Also note that the colour
factor for $qg\to qgg$ is ambiguous up to $1/N_C^2$ and that 
this variation is included in the MIN/MAX variation. 
\begin{table}[t]
\begin{center}
\begin{tabular}{lr|r|rrrrrr|rrr}
\hline 
& ${\cal C}$ & $C_{-1,-1}$ & $C_{-1,0}$ & $C_{0,-1}$ & $C_{-1,1}$ & $C_{1,-1}$ &
$C_{-1,2}$ & $C_{2,-1}$ &  $C_{0,0}$ & $C_{1,0}$ & $C_{0,1}$ \\
\hline
GGG\\
$q\bar{q}\to qg\bar{q}$ & $\frac83$ & 2 & -2 & -2 & 1 & 1 & 0 & 0 & 0
& 0 & 0
\\[1mm]
$qg\to qgg$& $\frac93$ & 2 & -2 & -2 & 1 & 1 & 0 & -1 & 2.5 & -1 & 1.5\\
\hline
MIN\\
$q\bar{q}\to qg\bar{q}$ & $\frac83$ & 2 & -2 & -2 & 1 & 1 & 0 & 0 & -6 & 4.5 & 4.5\\[1mm]
$qg\to qgg$& $\frac83$ & 2 & -2 & -2 & 1 & 1 & 0 & -1 & -8 & 8 & 7\\
\hline
MAX\\
$q\bar{q}\to qg\bar{q}$ & $\frac83$ & 2 & -2 & -2 & 1 & 1 & 0 & 0 & 2 & 1.5 & 1.5\\[1mm]
$qg\to qgg$& $\frac93$ & 2 & -2 & -2 & 1 & 1 & 0 & -1 & 2 & 1.5 & 1.5\\
\hline
\end{tabular}
\caption{Colour factors ${\cal C}$ (in the \textsc{Vincia}
  normalization \cite{Giele:2007di}) and Laurent coefficients
  $C_{\alpha,\beta}$ for the antennae used in
  this study. The
  coefficients with at least one negative index are universal (apart
  from a re-parametrization ambiguity for gluons). The positive-index
  coefficients are arbitrary and are here varied between MIN and MAX.
\label{tab:coefficients}}
\end{center}
\end{table}

\begin{figure}[tp]
\begin{center}
\begin{tabular}{cc}
\includegraphics*[scale=0.38]{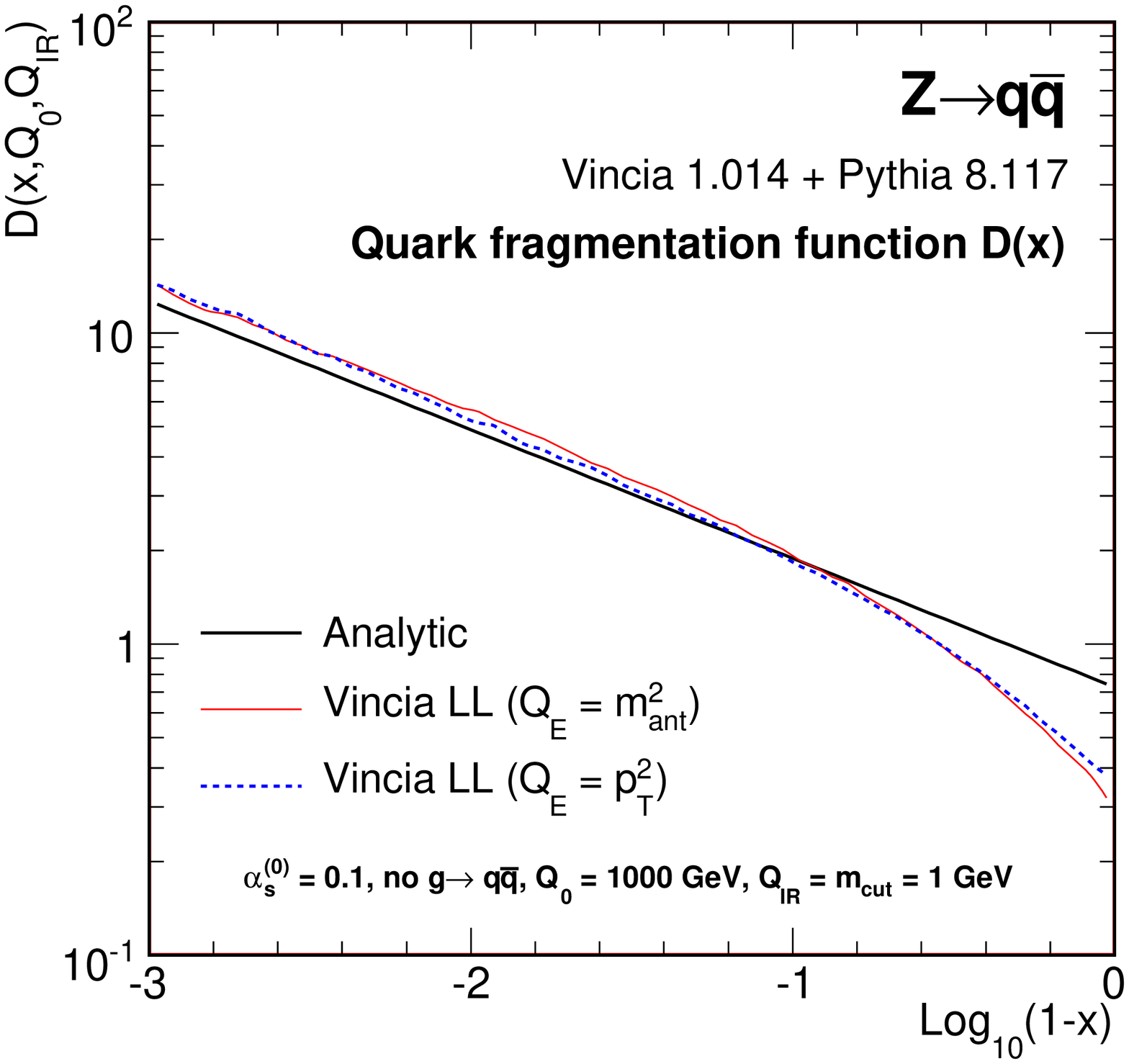} &
\includegraphics*[scale=0.38]{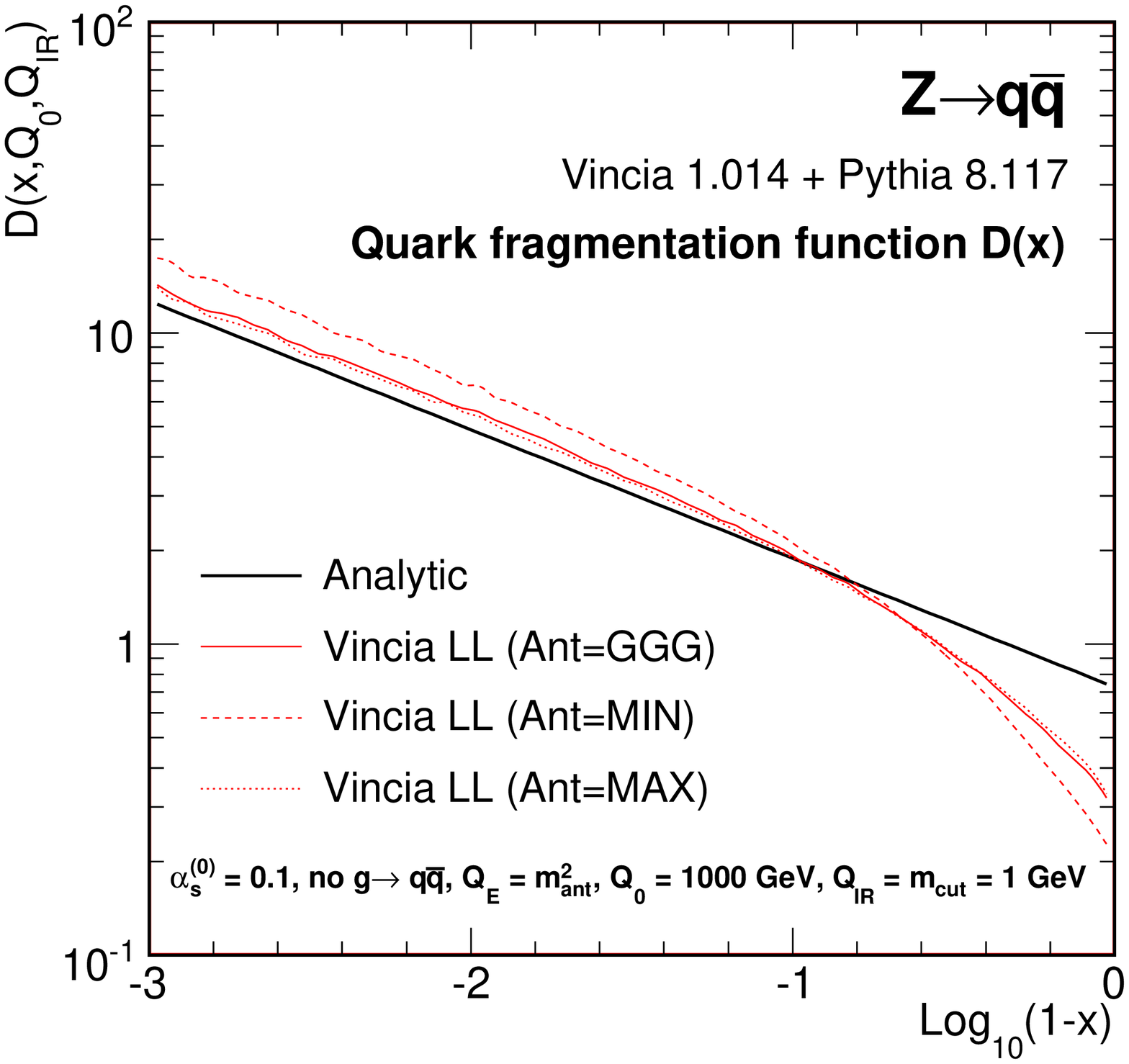} \\
\includegraphics*[scale=0.38]{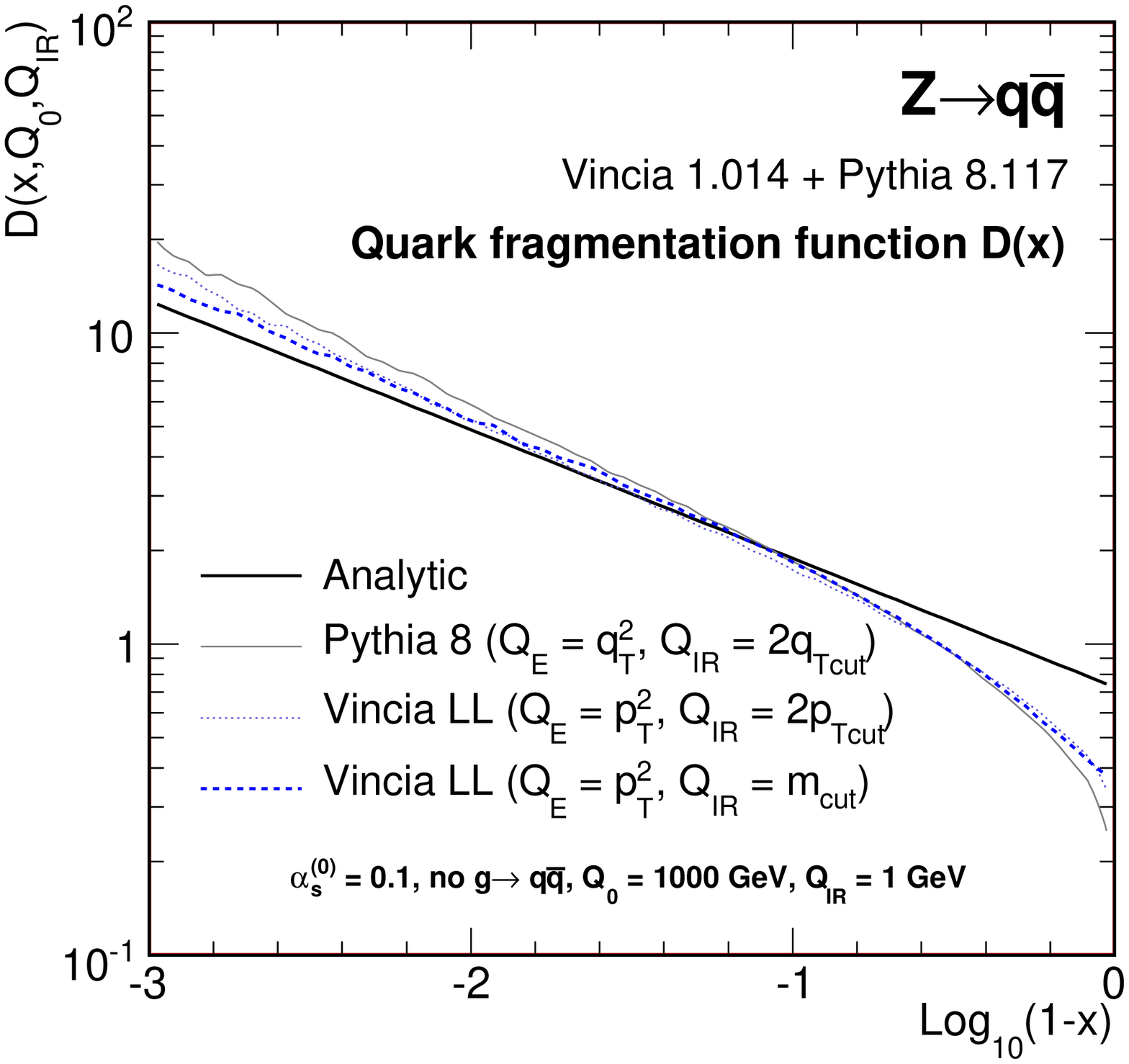} &
\includegraphics*[scale=0.38]{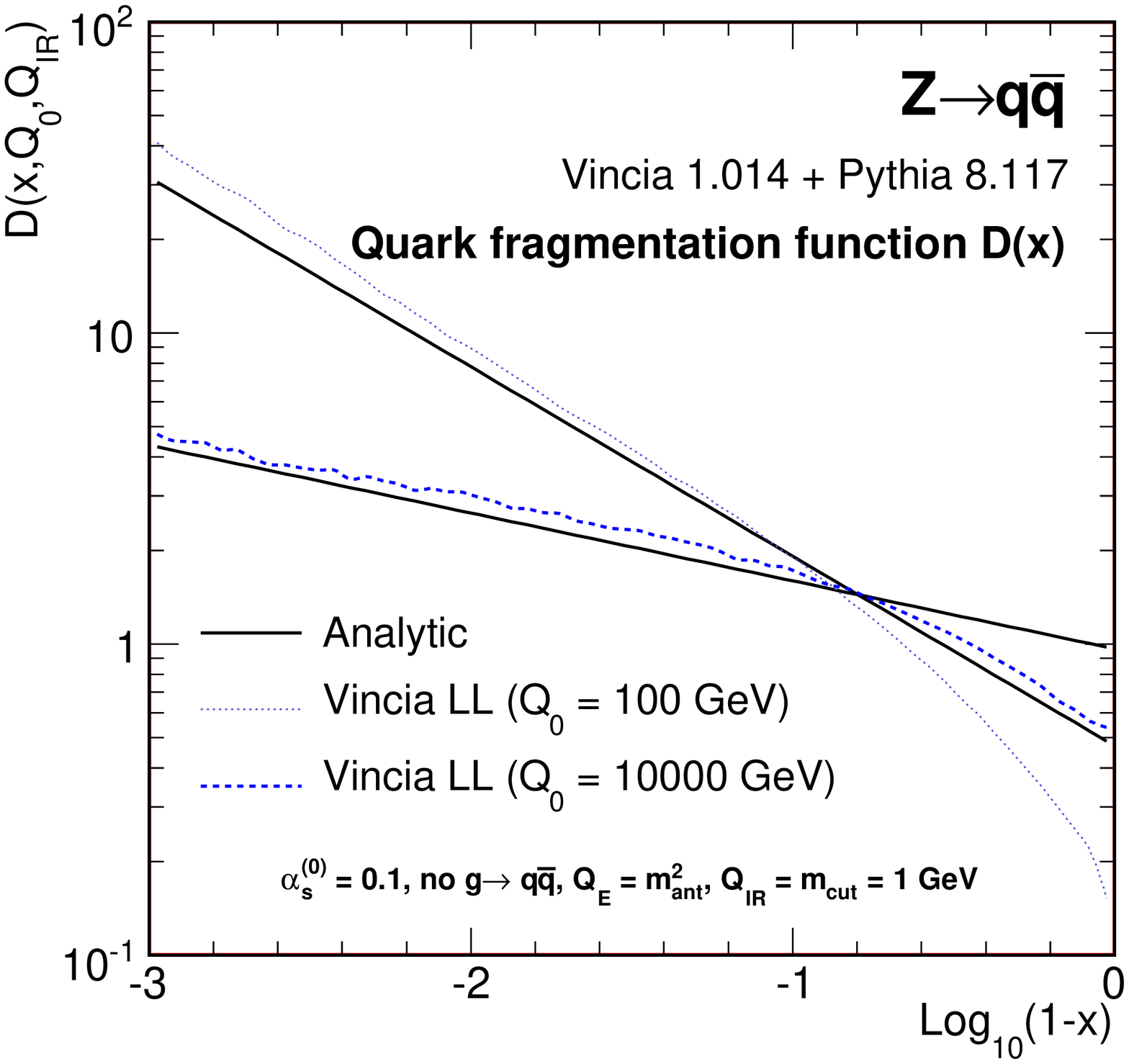}
\end{tabular}\vspace*{-1mm}
\caption{The quark fragmentation function as obtained with
  \textsc{Vincia} and \textsc{Pythia 8}, compared to the analytical
  expression, eq.~(\ref{eq:dAna}). {\sl Top Left:} variation of the
  evolution variable (\pT{}-ordering vs.~antenna-mass-ordering). 
  {\sl Top Right:} variation of the antenna
  functions (the Gehrmann-Glover functions vs.~the \textsc{Vincia} MIN
  and MAX variations). {\sl Bottom Left:} variation of the generator and of the
  hadronization cutoff contour (\textsc{Pythia 8} vs.~\pT{}-ordered
  \textsc{Vincia} with a cutoff in \pT{} or antenna-mass). 
	 {\sl Bottom Right:} variation of the
  starting scale ($m_Z = 10^2$ GeV vs.~$m_Z = 10^4$ GeV). 
  In all cases, $g\to q\bar{q}$ branchings were
  switched off, a constant $\alpha_s = 0.1$ was used, and $D$ was
  normalized to the number of inclusive 3-parton events at the cutoff.
\label{fig:DVincia}}
\end{center}
\end{figure}

In \textsc{Pythia 8}, the transverse-momentum variable, which we shall
 here call $q_T$ to distinguish it from the other definitions, agrees with the 
 \textsc{Ariadne} definition in the infrared limit, but differs
 from it by up to a factor of 2 away from that limit, see 
\cite{Sjostrand:2004ef}. The splitting functions in \textsc{Pythia} 8
 are the ordinary DGLAP ones, augmented by matching to the tree-level 
$Z\to  qg\qbar$ matrix element. 

In this study, we do not include variations of the kinematics
maps beyond that offered by the default \textsc{Pythia 8} 
and \textsc{Vincia} choices. \textsc{Pythia 8} 
uses a partitioned-dipole-like map in which a
``recoiler'' recoils longitudinally (in the dipole centre-of-mass
frame) against a ``radiator''. \textsc{Vincia} by default uses the
$gg\to ggg$ dipole-antenna map of \textsc{Ariadne} for all
branchings. In the dipole-antenna case, no special distinction is made
between radiator and recoiler; instead the proper collinear 
limiting behaviour is obtained by a rotation angle going to $0^\circ$
or $180^\circ$ in the respective limits. 

In fig.~\ref{fig:DVincia}, we show 4 plots illustrating the quark
fragmentation function with these variations, as compared to the
analytic expression, eq.~(\ref{eq:dAna}). The top left plot
illustrates that one obtains good agreement between \textsc{Vincia}
and the analytic expression, irrespective of the choice of evolution
variable. (Recall that the validity of the analytic expression is
limited to the region $\ln(1-x) \ll 0$.)  
It should be noted, though, that it could still be possible
to obtain bigger variations for more extreme variations of the choice
of evolution variable. 
The top right plot illustrates the dependence on the choice
of antenna functions. As expected, this variation is larger, since this
dependence enters already at first order in the shower
expansion. Nonetheless, the asymptotic slope of all the curves agrees
with the analytic expectation. In the lower left-hand plot, we show
the variation between \textsc{Pythia 8} and \textsc{Vincia}, with two
different choices for the hadronization cutoff for the latter, either
in $\pT{}$ or in antenna mass. Though we emphasize 
that this distribution is very infrared sensitive, the dependence on
the choice of hadronization cutoff here seems rather mild. The
\textsc{Pythia} 8 curve has a slightly different slope than the
\textsc{Vincia} ones, but still appears to be within the uncertainty
spanned by the variations above. Finally, in the lower right-hand plot,
we show the results for two alternative starting scales, one at 100
GeV and the other at 10000 GeV. As expected, the agreement improves
with increasing energy (or, more precisely, with increasing
$Q_0/Q_{\mathrm{IR}}$).  

In summary, we find no evidence in either partitioned-dipole or
dipole-antenna showers of a breakdown of agreement with DGLAP-based
predictions of the quark fragmentation function, 
provided that ``infrared sensible'' evolution variables are
chosen. The definition of infrared 
sensible is that both infinitely soft and collinear emissions should
be classified as unresolved for any finite value of the evolution
variable. 

\subsection{Comparison to second-order QCD}

A complementary check on the accuracy of the shower can be obtained by
comparing its second-order expansion to second-order QCD
matrix elements. This is simplest for the dipole-antenna shower, for
which the number of possible histories for each phase space point
grows less fast than for the partitioned-dipole case, so in this
subsection, we shall only use dipole-antenna showers for the
comparisons, but we emphasize that the results should be qualitatively
similar for the partitioned-dipole case. This part of our 
study is similar to a previous comparison of \textsc{Ariadne} to
second-order QCD by Andersson et al.~\cite{Andersson:1991he}. 

In phase space regions dominated by leading logs, the
ratio shower/matrix-element should be unity. In phase space regions
dominated by hard wide-angle emissions, the shower could in principle be
arbitrarily far from the matrix element, and finally in regions
dominated by subleading logs (such as regions with two emissions at
the same scale), the subleading-log properties of the shower can be
probed. 

To perform this test independently of the shower generator, 
we use \textsc{Rambo} to generate a large number of evenly distributed
4-parton phase space points. For each phase space point, we evaluate
the leading-colour 4-parton antenna function, 
\begin{equation}
A_{4\mathrm{LC}} = \frac{|M_{4\mathrm{LC}}(p_1,p_2,p_3,p_4)|^2}{|M_2(s)|^2}~, 
\end{equation}
as given by Gehrmann et al.~\cite{GehrmannDeRidder:2005cm} 
(counter checked with \cite{Ellis:1980wv} to protect against typos).  
 
We then compute the tree-level leading-colour LL antenna-shower approximation
corresponding to the same phase space point, based on nested $2\to 3$
branchings. For 4 partons, there are
two possible antenna-shower histories;
\begin{itemize}
\item {\sl A)} parton 2 emitted between partons 1 and 3.\\
 The 4-parton
  evolution scale is then $Q_{4A}^2 = Q_E^2(1,2,3)$. 
\item {\sl B)} parton 3 emitted between partons
2 and 4.\\The 4-parton evolution scale is then $Q_{4B}^2 = Q_E^2(2,3,4)$,
\end{itemize} 
with $Q_E$ denoting a generic evolution variable. We shall 
here consider energy-ordering, $\pT{}$-ordering, and antenna-mass
ordering. 

We note that a similar study for parton- or partitioned-dipole
(Catani-Seymour) showers would need to consider 8 possible paths from
2 to 4 partons \cite{Weinzierl:2003fx}: two possible radiators in
the first $2\to 3$ step (the quark and the antiquark), and 4 possible
radiators in the subsequent $3\to4$ step (treating the two ``sides''
of the gluon, which are generally associated with
different kinematics mappings, as separate). The dipole-antenna shower is
thus very economic in the number of terms generated at each successive
order. 

Using a new clustering algorithm that contains the
inverse of the \textsc{Vincia} $2\to3$ kinematics maps, we may perform 
clusterings of the type $(a,r,b)\to(\hat{a},\hat{b})$ in a way that
exactly reconstructs the intermediate 3-parton configurations that would have
been part of the shower history for each 4-parton 
test configuration, for each of
the paths A and B\footnote{Note that the inversion of 
\textsc{Vincia} by this clustering algorithm is exactly one-to-one, with no
approximation made. This was validated by reclustering a large
number of actual branchings generated by the shower and recovering
the pre-branching configurations exactly, including global
orientations, etc.}. This gives us an exact tree-level 
reconstruction of how the antenna shower would have populated each
path. We can now use this to test the shower approximation over all
of 4-parton phase space. 

We shall do this by plotting the ratio
\begin{equation}
R^0_{4} = \frac{
 A_{q\qbar}(\widehat{12},\widehat{23},4) A_{qg}(1,2,3) 
 + A_{q\qbar}(1,\widehat{23},\widehat{34}) A_{g\qbar}(2,3,4)  
}{A_4(1,2,3,4)}~,\label{eq:R4}
\end{equation}
with hatted variables $\widehat{ij}$ denoting clustered momenta. 
$R$ thus gives a direct measure of the amount of over- or under-counting by
the shower approximation, with values greater than unity corresponding
to over-counting and vice versa.

The ratio above, eq.~(\ref{eq:R4}), contains nested products of
antennae, identical to the subtraction terms that would be used in a
fixed-order calculation. This does not take into account the ordering
condition imposed in the shower, however. To impose this condition, we
must include step functions in the shower approximation, as follows
\begin{equation}
R^{E}_{4} = \frac{
 \Theta(Q_{3A}-Q_{4A}) A_{q\qbar}(\widehat{12},\widehat{23},4) A_{qg}(1,2,3) 
 + \Theta(Q_{3B}-Q_{4B}) 
A_{q\qbar}(1,\widehat{23},\widehat{34}) A_{g\qbar}(2,3,4) }{A_4(1,2,3,4)}~,
\label{eq:R4E}
\end{equation}
where 
\begin{equation}
\begin{array}{rclcrcl}
Q_{4A} & = & Q_E(1,2,3) & ; & Q_{3A} & = & Q_E(\widehat{12},\widehat{23},4)\\
Q_{4B} & = & Q_E(2,3,4) & ; & Q_{3B} & = & Q_E(1,\widehat{23},\widehat{34})
\end{array}~.\label{eq:4pAB}
\end{equation}
The ratio $R^E_4$ now faithfully reproduces the shower approximation
expanded to tree level, phase space point by phase space point, 
for an arbitrary choice of evolution variable,
$Q_E$. 

Since the full 4-parton phase space has more dimensions than can fit
on paper, and since the leading singularity of the gluon emission
antenna functions goes like $\pT{}^{-2}$ (with $\pT{}$ defined as 
in eq.~(\ref{eq:type1})), we project the full phase
space onto two $\pT{}$ values, one of which we choose to correspond 
to the initial $2\to 3$ step of a would-be shower history 
and the second to the $3\to4$ step. Specifically, the ordinate along
the $y$ axis will be
\begin{equation}
\mbox{$y$ axis : }\hspace*{1cm}\pT{;2}=\mathrm{min}(\pT{}(1,2,3),\pT{}(2,3,4))~,\hspace*{1cm}
\end{equation}
corresponding to the second branching, and the 
ordinate along the $x$ axis will be the $\pT{}$ value of the
reclustered 3-parton configuration corresponding to the
$\mrm{min}(...)$,
\begin{equation}
\mbox{$x$ axis : }\hspace*{1cm}
\pT{;1}=\left\{\begin{array}{ll}
\pT{}(\widehat{12},\widehat{23},4) & ~;~\pT{}(1,2,3) < \pT{}(2,3,4)\\[2mm]
\pT{}(1,\widehat{23},\widehat{34}) & ~;~ \pT{}(2,3,4) < \pT{}(1,2,3)\\
\end{array}\right.~.\hspace*{1cm}
\end{equation}

Each point in $(\pT{;1},\pT{;2})$-space thus contains a distribution of all
4-parton configurations with that particular combination of \pT{;1}
and \pT{;2} values. We shall plot both
the average of this distribution, which we call $<R_4^E>$, 
as well as a measure of the spread of the distribution, which we
define as
\begin{equation}
\mathrm{RMS}(R_4^E)~=~10^{\mathrm{RMS}(\log_{10}(R_4^E))}~ \ge~1~.
\end{equation}
By using this form we probe the average factor 
of deviation from unity rather than the absolute measure of
the deviation itself. (I.e., we want a point with $R_4^E=0.1$ to count
as having a deviation of a factor of ten, rather than an absolute 
deviation of 0.9, from unity.) A special case is when we encounter dead
zones in which the shower answer is zero, and hence the factor of
deviation would nominally be infinite. When computing the RMS above we
therefore put a floor on the deviation at a factor 0.01 times the
matrix element. 

\begin{figure}[tp]
\begin{center}
\begin{tabular}{ccc}
\plotset{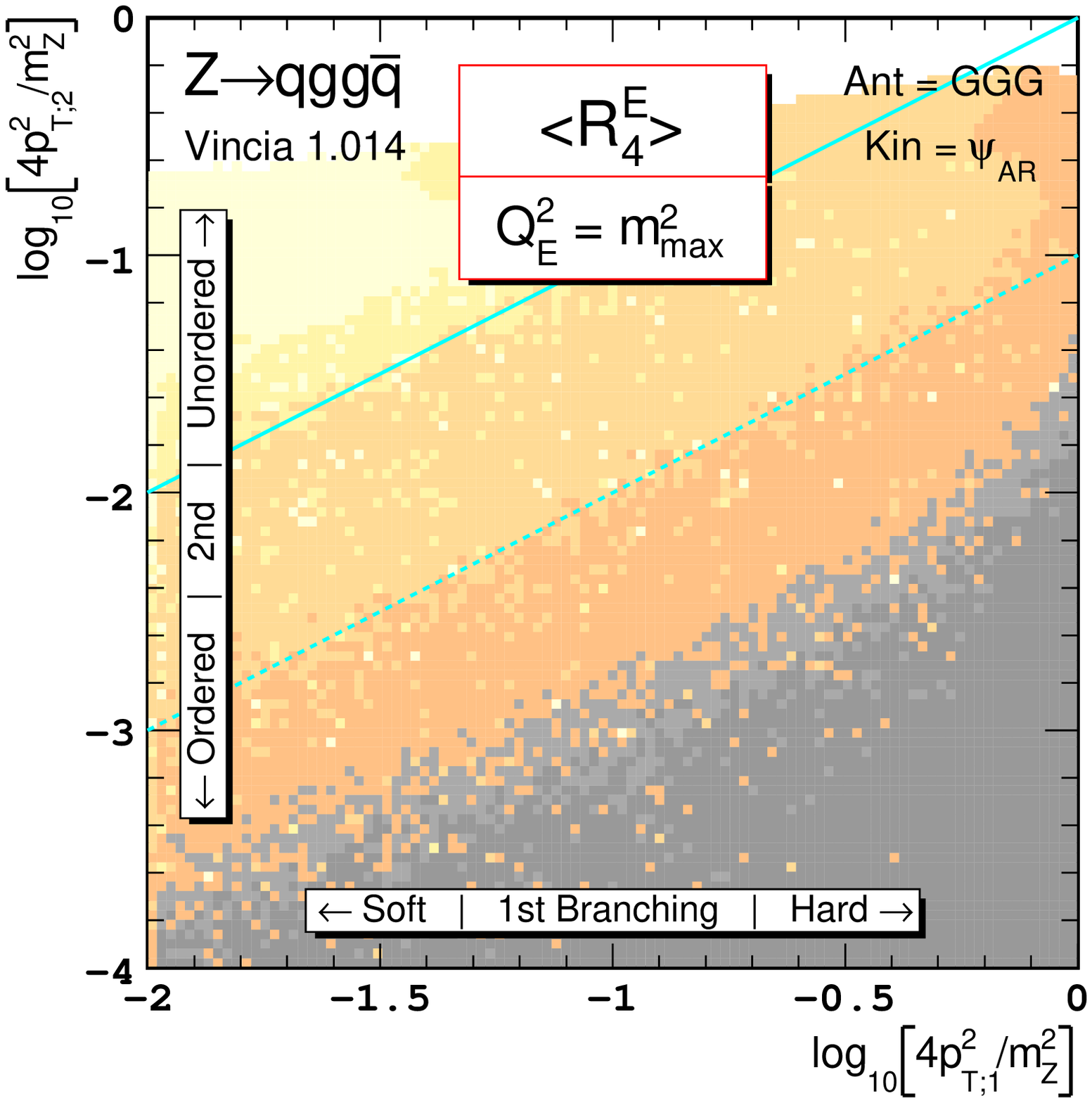}\\[-2.5mm]
\plotset{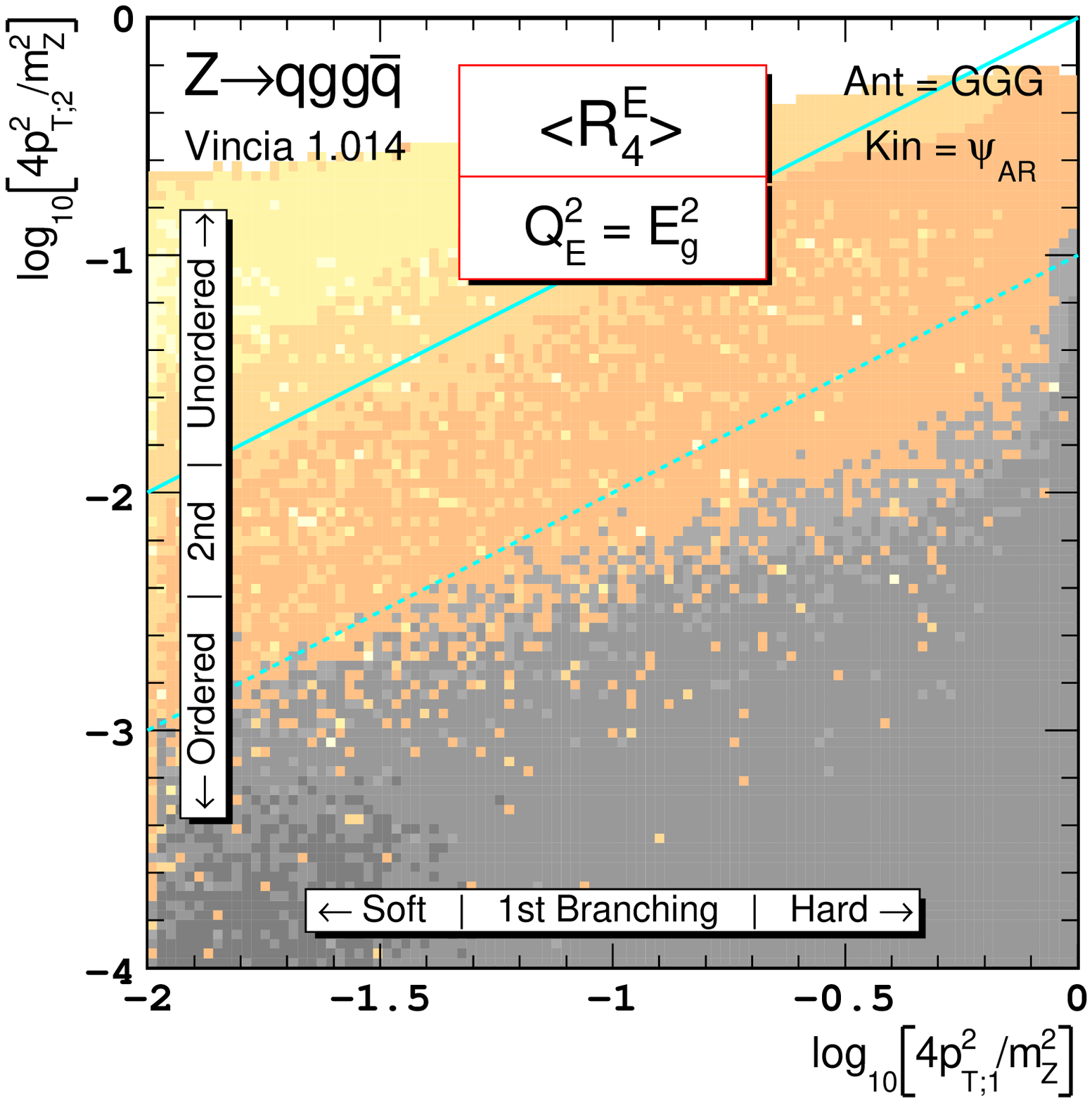}\\[-1mm]
\end{tabular}
\caption{{\sl Left:} Average of the $R_4^E$ 
  distributions, eqs.~(\ref{eq:R4}) and (\ref{eq:R4E}), 
  for no ordering ({\sl top}) and 
  energy-ordering ({\sl
    bottom}). Diagonal lines indicate boundaries between unordered,
  ordered, and strongly ordered regions (doubly strongly ordered
  region is at origo).  {\sl Right:}
  the RMS of the factor of deviation from unity of the $R_4^E$ distributions. 
  \label{fig:R4}}
\end{center}
\end{figure}

\begin{figure}[tp]
\begin{center}
\begin{tabular}{ccc}
\plotset{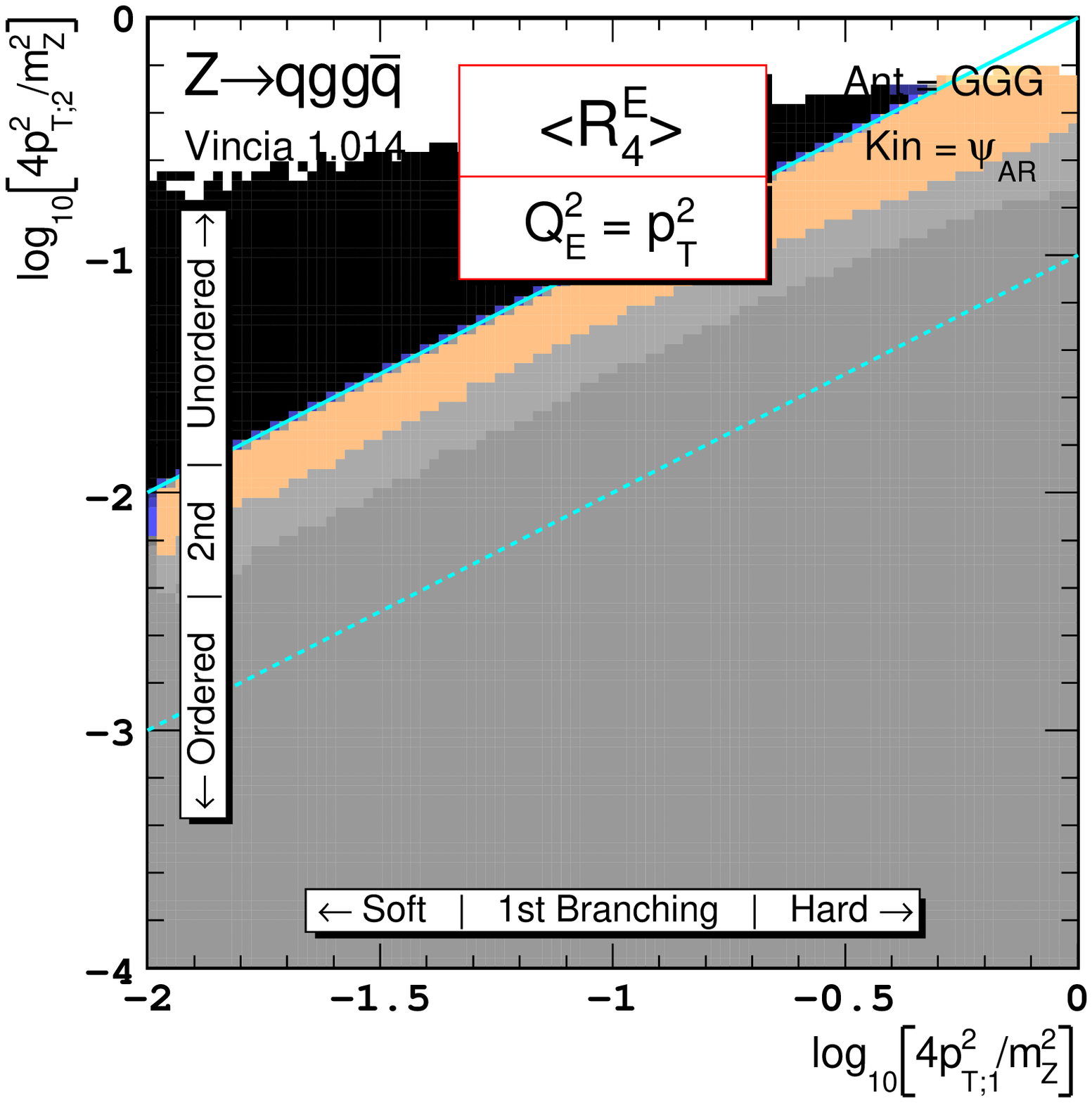}\\[-2.5mm]
\plotset{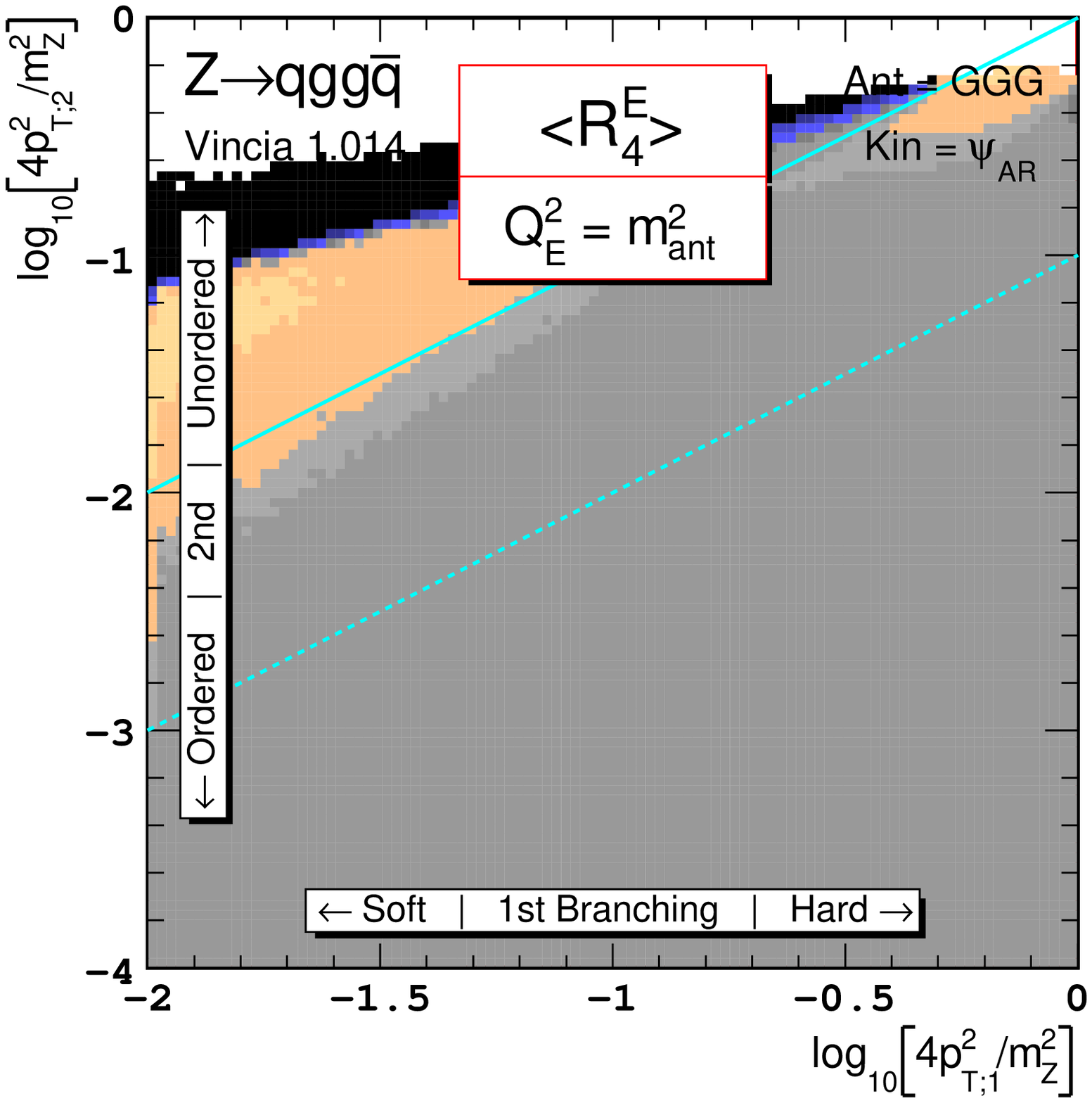}\\[-1mm]
\end{tabular}
\caption{{\sl Left:} Average of the $R_4^E$ 
  distributions, eq.~(\ref{eq:R4E}), 
  for $\pT{}$-ordering ({\sl
    top}) and antenna-mass-ordering ({\sl bottom}). 
  Diagonal lines indicate boundaries between unordered,
  ordered, and strongly ordered regions (doubly strongly ordered
  region is at origo).  {\sl Right:}
  the RMS of the factor of deviation from unity of the $R_4^E$
  distributions. The RMS distribution for antenna-mass-ordering (on
  the lower right) is somewhat affected by the occurrence of 
  a few dead points in a region extending towards the lower left,
  which ``artificially'' increase the RMS in that region. \label{fig:R4-2}}
\end{center}
\end{figure}

In figs.~\ref{fig:R4} and \ref{fig:R4-2}, we show 
the average and RMS for four different ordering variables; for
comparison, we first show the result \emph{without} any
ordering imposed, as in eq.~(\ref{eq:R4}), i.e., a simple product of nested 
antennae with no $\Theta$ functions imposed. This is equivalent to the
subtraction terms constructed for fixed-order calculations and can be
represented by ``ordering'' in the variable $m^2_{\mrm{max}} =
\mathrm{max}(s_{12},s_{23})$.
We then compare to the ordered results, eq.~(\ref{eq:R4E}),
for 
energy-ordering (as defined by Dokshitzer and Marchesini, i.e., 
ordering in the energy of the emitted parton in the CM of the $Z$
boson), \pT{}-ordering (as defined in eq.~(\ref{eq:type1})), and
antenna-mass-ordering (as defined in eq.~(\ref{eq:type2})). 
We may identify several regions of interest on the plots shown
in figs.~\ref{fig:R4} and \ref{fig:R4-2}:
\begin{itemize}
\item {\sl Origo:} 
  double-LL singular region: $\pT{;2} \ll \pT{;1} \ll s$, i.e., two
  widely separated jets plus two strongly ordered emissions. Should be
  correctly described by any LL shower.
\item {\sl Top right-hand corner:} 
  Hard region: $\pT{;1}\sim\pT{;2}\sim s$, i.e., 4 widely separated
  jets. Should be correctly described only by the 4-parton
  matrix element. 
\item {\sl Area below diagonal dashed (cyan) line (bottom right-hand
  corner):} region in which the
  second emission is strongly ordered with respect to the first:
  $\pT{;2} \ll \pT{;1}$. Should be correctly described by any LL shower matched to
  the 3-parton matrix element. 
\item {\sl Intersection of $y$-axis with diagonal solid (cyan) line:} 
  Single-NLL (double-emission) region:
  $\pT{;2}\sim\pT{;1}\ll s$, i.e. 2 widely separated jets plus one
  strongly ordered $2\to 4$ emission (two powers of
  $\alpha_s$ but only one large scale difference). Should  be correctly 
  described only by the 4-parton matrix element and/or by an NLL shower. 
\item {\sl Area above diagonal solid (cyan) line (except top right-hand
  corner, see above):} 
  \pT{}-unor\-dered
  region. Corresponds to a dead zone in a shower ordered in
  \pT{}. Although this zone occupies a relatively large area in our
  projection, this is chiefly an artifact of our choice of variables. 
  The actual phase space volume in this region amounts to 
  roughly 1.5\% of the full 4-parton phase space. 
\end{itemize}

Our operational definition of a ``correctly described region'' we
shall here take to be that both the average of the $R_4^E$
distribution as well as its RMS factor  in that region
should tend to unity.  If so, this means that the shower is not only getting
the average of the distribution right, but there are also no large
fluctuations on either side of the average. 
If only the average is unity but the RMS factor is not, then the
interpretation is that the shower is still over- and under-counting
individual phase space points, and hence the relevant part of phase
space is not being populated accurately.

In the top row of Fig.~\ref{fig:R4}, we show the average and RMS
factor without imposing any ordering condition at all (apart from that
implied by the nested phase spaces). The central grey colour towards
the bottom right indicates that the shower approximation deviates less
than 10\% from the matrix element there, and the two neighbouring grey shades
indicate 20\% deviation. The next colours (lighter red and darker blue shades) 
represent factors of 2, 5, and 10, respectively. The dominance of
light shades in the upper left half of the plots thus indicate that
the nested LL antenna products, without ordering, exhibit 
a large over-counting whenever the 2$^{\mathrm{nd}}$ emission
does not have a  \pT{} several orders of magnitude smaller than the
1$^\mrm{st}$ one. 

As expected, the situation for energy-ordering is actually worse (bottom plot
row). Though the agreement improves slightly for hard 2$^\mrm{nd}$ emissions, 
we see a disturbing trend towards systematic
undercounting in the double-LL region, indicated by the darkening
shades towards origo. What is even more disturbing is that the RMS
factor does not follow the average, but instead blows up towards the
LL singular regions. This means that the ``good'' average agreement on
the left-hand side is really obtained by cancellations between
large over- and undercounting of individual phase space
points. The desired LL singular behaviour is therefore observed not to
be obtained for this evolution variable. 

Finally, the quite impressive properties of \pT{}-ordered dipole-antenna showers 
are evident in the near-unity average value of both $R^E_4$ and its
RMS factor of deviation over all LL-dominated 
phase space regions in the top left-hand plot of fig.~\ref{fig:R4-2}, 
as was also noted in the study of  Andersson et
al.~\cite{Andersson:1991he}. Though we 
here use a different set of antenna functions and kinematics maps, 
we see that the excellent basic agreement with the second-order QCD matrix
elements is retrieved also in our case.  
We note that by definition a \pT{}-ordered shower does not generate any points
in the \pT{}-unordered region above the diagonal (cyan) solid line.
(Again, these dead zones were also
pointed out in the study by Andersson et
al.~\cite{Andersson:1991he}). 
As mentioned, we find that roughly 1.5\% of the full 4-parton phase space is left
unpopulated by this particular ordering variable
(antenna-mass-ordering gives a similar number). Keep 
in mind that this is 4-parton space though; none of these showers have
any dead zones in $3$-parton phase space. 

The case of
antenna-mass-ordering (fig.~\ref{fig:R4-2}, bottom row) is similar to
$\pT{}$-ordering, but its dead zones do not follow strict
contours of $\pT{}$, and hence 
the RMS factor looks ``artificially'' large over that part of phase
space in which dead points exist, roughly the area above the diagonal
of the plot. 

We plan to return to the issue of dead zones in a future paper, but
note that since they are located in the NLL-dominated region,
they do not affect the conclusions we wish to make here concerning the
LL behaviour of the evolution choices. We thus 
restrict ourselves to the conclusion  
that $\pT{}$-ordered dipole-antenna showers appear to give an 
excellent approximation to 
the full 4-parton matrix element over all LL dominated regions of
phase space (below the diagonal (cyan) dashed line). When
\pT{}-ordering is imposed, the RMS factor of deviation furthermore registers an
impressive sharpening-up of the $R^E_4$ distribution, yielding much
larger regions of unity RMS factor than the corresponding case without
ordering. Also antenna-mass-ordering represents a substantial
improvement, although the improvement in the RMS is slightly masked by
the fact that our projection ``smears out'' its dead zones over a
larger area of the plot than for \pT{}-ordering. 

Energy ordering,
on the other hand, effectively introduces artificially under-counted zones in 
the doubly-LL singular region, while still not removing the over-counting
that was already present in the same region without ordering --- hence the
RMS measure of deviation actually worsens as we go further into the
singular region. 
The factorization implied by this choice of evolution
variable is thus clearly not consistent with the structure of QCD.


\section{Conclusions}
\label{sect:conclusions}

In this paper we studied shower algorithms based on partitioned-dipoles
and dipole-antennae.
In particular we investigated the behaviour in the collinear limit and showed that
with an ``infrared-sensible'' definition of the evolution variable they
reproduce the DGLAP evolution equation. The definition of infrared
sensible is that both infinitely soft and collinear emissions should
be classified as unresolved for any finite value of the evolution
variable. Examples of such choices are $k_\perp$-ordering or
mass-ordering (ordering in virtuality for partitioned-dipole and
antenna-mass for dipole-antenna showers, respectively). 
On the other hand, ordering in the energy of the emitted particle is not infrared sensible (it
classifies infinitely collinear emissions as being resolved) and does
not reproduce the DGLAP equation. 

In addition to these analytic arguments, we have also presented 
a numerical study, making use of existing dipole shower algorithms.
We demonstrated that the DGLAP behaviour of the quark
fragmentation function is reproduced by these models for a range of
different infrared sensible shower algorithms, in particular
\pT{}-ordered ones. In addition we compared dipole-antennae to
second-order QCD matrix elements and again retrieve good agreement in
the strongly ordered (LL-dominated) region for \pT{}-ordering, but not for
energy-ordering. 

\subsection*{Acknowledgements}

We would like to thank G. Gustafson, G. Marchesini, Z. Nagy and D. Soper
for useful discussions. Special thanks go to H. Jung for organizing
a discussion meeting at DESY.

\bibliography{biblio}
\bibliographystyle{h-physrev3}

\end{document}